\documentclass[journal]{IEEEtran} 
\usepackage{graphicx,psfrag}
\usepackage[cmex10]{amsmath}
\usepackage{amsfonts,amssymb,latexsym,hyperref}
\usepackage[nospace,nocompress]{cite}
\usepackage[usenames,dvipsnames]{color}
\usepackage{algorithm,algorithmic,setspace}
\usepackage{multirow}
\usepackage{subcaption}
\allowdisplaybreaks

\makeatletter
\def\BState{\State\hskip-\ALG@thistlm}
\makeatother

 \renewcommand{\tilde}{\widetilde}
 \renewcommand{\hat}{\widehat}
 \renewcommand{\bar}{\overline}
 \newcommand{\defn}{\triangleq}

 \newcommand{\tvec}[1]{\ensuremath{\Tilde{\boldsymbol{#1}}}}
 \newcommand{\ovec}[1]{\ensuremath{\Bar{\boldsymbol{#1}}}}
 \newcommand{\hvec}[1]{\ensuremath{\Hat{\boldsymbol{#1}}}}
    
 \renewcommand{\vec}[1]{\ensuremath{\boldsymbol{#1}}}
 \newcommand{\mat}[1]{\ensuremath{\begin{bmatrix}#1\end{bmatrix}}}

 \newcommand{\norm}[1]{\ensuremath{\| #1 \|}}
 \newcommand{\mc}[1]{\ensuremath{\mathcal{#1}}}

 \newcommand{\Real}{{\mathbb{R}}}
 \newcommand{\Complex}{{\mathbb{C}}}

 \newcommand{\tran}{^{\textsf{T}}}
 
 \newcommand{\giv}{\,|\,}
 \newcommand{\biggiv}{\,\big|\,}

 \DeclareMathOperator{\real}{Re}
 \DeclareMathOperator{\imag}{Im}

 \DeclareMathOperator{\E}{\mathbb{E}}
 \DeclareMathOperator{\var}{var}
 \DeclareMathOperator{\cov}{Cov}

 \DeclareMathOperator{\tr}{tr}
 \DeclareMathOperator{\diag}{diag}
 
 \DeclareMathOperator{\Diag}{Diag}

 \DeclareMathOperator*{\argmin}{arg\,min}
 \DeclareMathOperator*{\argmax}{arg\,max}

 \renewcommand{\eqref}[1]{(\ref{eq:#1})}
 
 \newcommand{\Figref}[1]{Figure~\ref{fig:#1}}
 
 \newcommand{\figref}[1]{Fig.~\ref{fig:#1}}

 \newcommand{\secref}[1]{Section~\ref{sec:#1}}
 \newcommand{\Secref}[1]{Section~\ref{sec:#1}}

 \newcommand{\algref}[1]{Algorithm~\ref{alg:#1}}
 \newcommand{\lineref}[1]{line~\ref{line:#1}}

\newcommand{\textb}[1]{\textcolor{black}{#1}}
\newcommand{\blue}{\color{black}}
\newcommand{\black}{\color{black}}

 \newcommand{\jj}{\mathrm{j}}
 \newcommand*\dif{\mathop{}\!\mathrm{d}} % differential in integrals
 \newcommand{\notk}{_{\setminus k}}

 \newcommand{\scalek}{\beta_k}
 \newcommand{\scalel}{\beta_l}
 
 \newcommand{\scalell}{\beta_l^2}

\newcommand{\rvecc}{\textbf{\textsf{c}}}

\newcommand{\ry}{\textsf{y}}

\newcommand{\rvz}{\textsf{z}}
\newcommand{\rvecz}{\textsf{\textbf{z}}}

\newcommand{\rvecp}{\textsf{\textbf{p}}}
\newcommand{\rvecs}{\textsf{\textbf{s}}}
\newcommand{\Qz}{\vec{Q}^{\rvecz}}
\newcommand{\Qp}{\vec{Q}^{\rvecp}}

\newcommand{\qz}{\vec{q}^{\rvecz}}
\newcommand{\qp}{\vec{q}^{\rvecp}}
\newcommand{\qps}{q^{\textsf{p}}}
\newcommand{\qs}{\vec{q}^{\rvecs}}
\newcommand{\qzs}{q^{\textsf{z}}}
\newcommand{\qns}{q^{\textsf{n}}}

\newcommand{\rvecr}{\textbf{\textsf{r}}}

\newcommand{\Qr}{\vec{Q}^{\rvecr}}

\newcommand{\qr}{\vec{q}^{\rvecr}}
\newcommand{\qc}{\vec{q}^{\rvecc}}

\newcommand{\pyz}{p_{\ry|\rvecz}}
\newcommand{\pc}{p_{\rvecc}}

\newcommand{\pcr}{p_{\rvecc|\rvecr}}

\newcommand{\pzyp}{p_{\rvecz|\ry,\rvecp}}

\newcommand{\MAPk}{_{k,\textsf{MAP}}}

\newcommand{\labelsize}{.75}

 \newfloat{myalgo}{tbhp}{mya}
  {\begin{myalgo}[#1]
        \centering
        \begin{minipage}{#2}
        \begin{algorithm}[H]}
  {\end{algorithm}
        \end{minipage}
        \end{myalgo}}
\begin{document}
\setlength{\arraycolsep}{0.8mm}

\title{Sketched Clustering via Hybrid Approximate Message Passing}
 
\author{
      Evan Byrne,\IEEEauthorrefmark{1} 
   Antoine Chatalic,\IEEEauthorrefmark{2}    
   R\'emi Gribonval\IEEEauthorrefmark{2}, \emph{IEEE Fellow},
   and Philip Schniter\IEEEauthorrefmark{1}, \emph{IEEE Fellow}%
	\thanks{\IEEEauthorrefmark{1}E. Byrne (byrne.133@osu.edu) and P. Schniter (schniter.1@osu.edu) are with the Department of Electrical and Computer Engineering at The Ohio State University, Columbus, OH, USA.}%
        \thanks{\IEEEauthorrefmark{2}A. Chatalic (antoine.chatalic@irisa.fr) and R. Gribonval (remi.gribonval@inria.fr) are with Univ. Rennes, Inria, CNRS, IRISA, France. A.~Chatalic received a travel grant from the French research network GdR MIA to visit The Ohio State University.}%
	\thanks{Please direct all correspondence to Philip Schniter, Dept.\ ECE, 2015 Neil Ave., Columbus OH 43210, phone 614.247.6488, fax 614.292.7596.}% 
        \thanks{\IEEEauthorrefmark{1}E. Byrne and P. Schniter acknowledge support from NSF grant 1716388 and MIT Lincoln Labs.}%
        \thanks{Portions of this work were presented at the 2017 Asilomar Conference of Signals, Systems, and Computers.}
   %        \IEEEauthorrefmark{2}%
%        Univ Rennes, Inria, CNRS, IRISA, France.
%        (antoine.chatalic@ens-rennes.fr, remi.gribonval@inria.fr)%

 }% 

\date{\today}
\maketitle

\begin{abstract}
In sketched clustering, a dataset of $T$ samples is first sketched down to a vector of modest size, from which the centroids are subsequently extracted.
Advantages include i) reduced storage complexity and ii) centroid extraction complexity independent of $T$.
For the sketching methodology recently proposed by Keriven et al., which can be interpreted as a random sampling of the empirical characteristic function, we propose a sketched clustering algorithm based on approximate message passing.
Numerical experiments suggest that our approach is more efficient than the state-of-the-art sketched clustering algorithm ``CL-OMPR'' (in both computational and sample complexity) and more efficient than k-means++ when $T$ is large.
\end{abstract}


%-------------------------------------------------------------------------------
\section{Introduction}              \label{sec:intro}

Given a dataset $\vec{X} \defn [\vec{x}_1,\dots,\vec{x}_T] \in \Real^{N\times T}$ comprising $T$ samples of dimension $N$, 
the standard clustering problem is to find $K$ centroids $\vec{C} \defn [\vec{c}_1,\dots,\vec{c}_K]\in\Real^{N\times K}$ that minimize the sum of squared errors (SSE)
\begin{equation}
\label{eq:sse}
\text{SSE}(\vec{X},\vec{C}) 
\defn \frac{1}{T} \sum_{t=1}^T \min_k \norm{\vec{x}_t-\vec{c}_k}_2^2.
\end{equation}
Finding the optimal $\vec{C}$ is an NP-hard problem \cite{Drineas:ML:04}.
Thus, many heuristic approaches have been proposed, such as the \emph{k-means} algorithm \cite{Steinhaus:BAPS:56,Jain:PRL:10}. 
%\cite{Steinhaus:BAPS:56,Jain:PRL:10}. 
Because k-means can get trapped in bad local minima, robust variants have been proposed, such as
\emph{k-means++} \cite{Arthur:SODA:07}, which uses a careful random initialization procedure to yield solutions with SSE that have on average $\leq 8(\ln K+2)$ times the minimal SSE.
The computational complexity of k-means++ scales as $O(TKNI)$, with $I$ the number of iterations, which is impractical when $T$ is large.

\subsection{Sketched Clustering}

In \emph{sketched clustering} \cite{Keriven:II:17,Keriven:ICASSP:17,Gribonval:17},
the dataset $\vec{X}$ is first sketched down to a vector $\vec{y}$ with $M = O(KN)$ components, from which the centroids $\vec{C}$ are subsequently extracted.
In the typical case that $K\ll T$, the sketch %$\vec{y}$ 
consumes much less memory than the original dataset. % $\{\vec{x}_t\}_{t=1}^T$.
If the sketch can be performed efficiently, then---since the complexity of centroid-extraction is invariant to $T$---sketched clustering may be more efficient than direct clustering methods when $T$ is large.
Note, for example, that k-means++ processes the $T$ data samples in $\vec{X}$ at every iteration, whereas sketched clustering processes the $T$ data samples in $\vec{X}$ only once, during the sketching step.

In this work, we focus on sketches of the type
proposed by Keriven et al.\ in \cite{Keriven:II:17,Keriven:ICASSP:17}, which use
$\vec{y}=[y_1,\dots,y_M]\tran$ with
\begin{equation}
y_m = \frac{1}{T}\sum_{t=1}^T \exp(\jj\vec{w}_m\tran\vec{x}_t) 
\label{eq:sk}
\end{equation}
and randomly\footnote{\textb{In \cite{Keriven:II:17} it was proposed to generate $\{\vec{w}_m\}$ as independent draws from a distribution for which $\vec{w}_m/\|\vec{w}_m\|$ is uniformly distributed on the unit sphere but $\|\vec{w}_m\|$ has a prescribed density.  More details are given in 
\secref{cl}.
%\secref{freq_gen}.
}}
generated $\vec{W} \defn [\vec{w}_1,\dots,\vec{w}_M]\tran \in\Real^{M\times N}$.
Note that $y_m$ in \eqref{sk} can be interpreted as a sample of the empirical characteristic function \cite{Feuerverger:AS:77}, i.e., 
%the characteristic function of a random vector with density $p(\vec{x})$,
\begin{equation}
\phi(\vec{w}_m) 
=\int_{\Real^N} p(\vec{x}) \exp(\jj \vec{w}_m\tran \vec{x}) \dif\vec{x} 
\end{equation}
under the empirical distribution
$p(\vec{x}) = \frac{1}{T}\sum_{t=1}^T \delta(\vec{x} - \vec{x}_t)$,
with Dirac $\delta(\cdot)$.
Here, each $\vec{w}_m$ can be interpreted as a multidimensional frequency sample.
The process of sketching $\vec{X}$ down to $\vec{y}$ via \eqref{sk} costs $O(TMN)$ operations, but it can be performed efficiently in an online and/or distributed manner. 

To recover the centroids $\vec{C}$ from $\vec{y}$, the state-of-the-art algorithm is \emph{compressed learning via orthogonal matching pursuit with replacement} (CL-OMPR) \cite{Keriven:II:17,Keriven:ICASSP:17}. 
It aims to solve 
\begin{align}
%\{\hvec{C},\hvec{\alpha}\} = 
\argmin_{\vec{C}} \min_{\vec{\alpha}: \vec{1}\tran\vec{\alpha}=1}\sum_{m=1}^M \bigg| y_m -\sum_{k=1}^K \alpha_k \exp(\jj \vec{w}_m\tran \vec{c}_k)  \bigg|^2
\label{eq:sk_clompr}
\end{align}
using a greedy heuristic inspired by the \emph{orthogonal matching pursuit} (OMP) algorithm \cite{PRK93} popular in compressed sensing.
With sketch length $M\geq 10KN$, CL-OMPR typically recovers centroids of similar or better quality to those attained with k-means++. 
%despite the lack of a direct link between the problem formulations \eqref{sse} and \eqref{sk_clompr}.
One may wonder, however, whether it is possible to recover accurate centroids with sketch lengths closer to the counting bound $M=KN$.
Also, since CL-OMPR's computational complexity is $O(MNK^2)$,
one may wonder whether it is possible to recover accurate centroids with computational complexity $O(MNK)$.

\subsection{Contributions}

To recover the centroids $\vec{C}$ from a sketch $\vec{y}$ of the form in \eqref{sk},
we propose the \emph{compressive learning via approximate message passing} (CL-AMP) algorithm, with computational complexity $O(MNK)$.
\textb{Numerical experiments show that, in most cases, CL-AMP accurately recovers centroids from sketches of length $M\geq 2KN$. 
This is an improvement over CL-OMPR, which typically requires $M\geq 10KN$. 
Our experiments establish these behaviors over many combinations of $K\in[5,50]$, $N\in[10,300]$, and sample numbers $T\in[10^5,10^8]$.} 
%which is an improvement over CL-OMPR's usual requirement that $M=10KN$.
%\textr{which makes it faster than CL-OMPR for large enough $K$.}
Experiments also show that CL-AMP recovers centroids faster and more accurately than k-means++ \textb{when $T$ is large, e.g., $T\geq 10^7$ in our numerical experiments}.
%CL-AMP can be understood as an application of the \emph{simplified hybrid generalized approximate message passing} (SHyGAMP) framework \cite{Byrne:TSP:16} to sketched clustering.
%SHyGAMP is a simplification of the HyGAMP from \cite{Rangan:TSP:17}, which is in turn a generalization of the GAMP approach from \cite{Rangan:ISIT:11}.
%Further details will be provided in the sequel.

We proposed a simple incarnation of the CL-AMP algorithm in the conference paper \cite{Byrne:ASIL:17}, with derivation details omitted due to space limitations.
In this paper, we present the full derivation of CL-AMP with an improved initialization and hyperparameter tuning scheme, and a much more comprehensive set of numerical experiments.
%(e.g., MNIST and frequency-estimation results).

The remainder of the paper is organized as follows.
In \secref{clamp}, we derive CL-AMP after reviewing relevant background on approximate message passing (AMP) algorithms. 
In \secref{num}, we present numerical experiments using synthetic and MNIST data, and we apply CL-AMP to multidimensional frequency estimation.
In \secref{conc}, we conclude.

%-------------------------------------------------------------------------------
\section{Compressive Learning via AMP} \label{sec:clamp}

\subsection{High-Dimensional Inference Framework} \label{sec:cl}

CL-AMP treats centroid recovery as a high-dimensional inference problem rather than an optimization problem like minimizing \eqref{sse} or \eqref{sk_clompr}.
In particular, it models the data using a Gaussian mixture model (GMM)  
\begin{align}
\vec{x}_t \sim \sum_{k=1}^K\alpha_k\mc{N}(\vec{c}_k, \vec{\Phi}_k),
\label{eq:gm} 
\end{align}
where the centroids $\vec{c}_k$ act as the GMM means, and the GMM weights $\alpha_k$ and covariance \textb{matrices} $\vec{\Phi}_k$ are treated as unknown parameters.
\textb{That is, $\{\vec{x}_t\}_{t=1}^T$ are assumed to be drawn i.i.d.\ from the GMM distribution \eqref{gm}.}
To recover the centroids $\vec{C}\defn[\vec{c}_1,\dots,\vec{c}_K]$ from $\vec{y}$, CL-AMP computes an approximation to the MMSE estimate
\begin{align}
\hvec{C} 
= \E \{ \vec{C} \giv \vec{y} \}  
\label{eq:exp_problem} ,
\end{align}
where the expectation is taken over the posterior density 
\begin{align}
p(\vec{C}|\vec{y})
&\propto p(\vec{y}|\vec{C}) p(\vec{C}) 
\label{eq:post} .
\end{align}
%and where $(\hvec{\alpha}$ and $\hvec{\Phi})$ are approximate maximum-likelihood estimates of $\vec{\alpha}\defn[\alpha_1,\dots,\alpha_K]\tran$ and $\vec{\Phi}\defn[\vec{\Phi}_1,\dots,\vec{\Phi}_K]$. 
In \eqref{post}, $p(\vec{y}|\vec{C})$ is the likelihood function of $\vec{C}$, and $p(\vec{C})$ is the prior density on $\vec{C}$.
The dependence of $p(\vec{y}|\vec{C})$ on $\{\alpha_k\}$ and $\{\vec{\Phi}_k\}$ will be detailed in the sequel.

\textb{As we now establish,}
the form of the sketch in \eqref{sk} implies that,
conditioned on the centroids $\vec{C}$ and the frequencies $\vec{W}$, the elements of $\vec{y}$ can be treated as i.i.d.
In other words, the sketch $\vec{y}$ follows a generalized linear model (GLM) \cite{McCullagh:Book:89}.
\textb{To establish this result, let us first} define the normalized frequency vectors 
\begin{align}
\vec{a}_m\defn\vec{w}_m/g_m \text{~with~} g_m\defn\|\vec{w}_m\|
\label{eq:am}
\end{align}
and the (normalized) transform outputs
\begin{align}
\vec{z}_m\tran
&\defn \vec{a}_m\tran\vec{C} \in \Real^K
\label{eq:zm} .
\end{align}
Then $p(\vec{y}|\vec{C})$ takes the form of a GLM, i.e.,
\begin{align}
p(\vec{y}|\vec{C}) 
&= \prod_{m=1}^M \pyz(y_m|\vec{a}_m\tran\vec{C}) 
\label{eq:glm},
\end{align}
for a conditional pdf $\pyz$ that will be detailed in the sequel.
%AMP methods have been successfully applied to many instances of GLM inference with large random transforms $\vec{A}$, such as quantized compressive sensing \cite{Kamilov:TSP:12}, binary classification \cite{Ziniel:TSP:15}, phase retrieval \cite{Schniter:TSP:15}, and multinomial logistic regression \cite{Byrne:TSP:16}.
%Thus, there is reason to believe that AMP methods may be applicable to the compressive learning problem \eqref{exp_problem}.

%We now derive the conditional pdf $\pyz$ in \eqref{glm} and assess its dependence on the parameters $\{\alpha_k\}$ and $\{\vec{\Phi}_k\}$ in \eqref{gm}.
From \eqref{sk} and the definitions of $\vec{a}_m$ and $g_m$ in \eqref{am}, we have 
\begin{align}
y_m 
&= \frac{1}{T}\sum_{t=1}^T \exp(\jj\vec{w}_m\tran\vec{x}_t) \\
&\approx \E\big\{\exp(\jj\vec{w}_m\tran\vec{x}_t) \textb{ \,\big|\,\vec{w}_m} \big\} 
\label{eq:ym1} \\
&= \sum_{k=1}^K \alpha_k \exp\bigg(\jj g_m \underbrace{\vec{a}_m\tran\vec{c}_k}_{\displaystyle \defn z_{mk}} - \frac{g_m^2}{2} \underbrace{\vec{a}_m\tran\vec{\Phi}_k \vec{a}_m}_{\displaystyle \defn \tau_{mk}} \bigg)
\label{eq:ym2} ,
\end{align}
where \eqref{ym1} holds under large $T$ and
\eqref{ym2} follows from the fact 
\begin{align}
\vec{w}_m\tran\vec{x}_t \textb{\,\big|\,\vec{w}_m}
&\sim \sum_{k=1}^K \alpha_k \mc{N}(g_m z_{mk},g_m^2\tau_{mk})
\end{align}
under \eqref{gm}, 
and the following well-known result \cite[p.153]{Papoulis:Book:91}:
\begin{align}
\E\{e^{\jj x}\} 
&= \exp\left( \jj \mu - \sigma^2/2 \right) \text{~when~} x\sim\mc{N}(\mu,\sigma^2) .
\end{align}
For $\vec{a}_m$ distributed uniformly on the sphere, the elements $\{\tau_{mk}\}_{m=1}^M$ in \eqref{ym2} concentrate as $N\rightarrow\infty$ \cite{Rudelson:ECP:13}, in that
\begin{align}
\tau_{mk} \stackrel{p}{\rightarrow} \E\{\tau_{mk}\} = \tr(\vec{\Phi}_k)/N \defn \tau_k
\label{eq:trRx} ,
\end{align}
as long as the peak-to-average eigenvalue ratio of $\vec{\Phi}_k$ remains bounded.
Thus, for large $T$ and $N$, \eqref{ym2} and \eqref{trRx} imply that
\begin{align}
y_m 
&= \sum_{k=1}^K \alpha_k \exp\bigg(\jj g_m z_{mk} - \frac{g_m^2 \tau_{k}}{2}\bigg)
\label{eq:ym3} ,
\end{align}
\textb{which implies that the inference problem depends on the covariance matrices $\{\vec{\Phi}_k\}$ only through the hyperparameters $\{\tau_k\}$.
Equation \eqref{ym3} can then} be rephrased as
\begin{align}
\pyz(y_m|\vec{z}_m;\vec{\alpha},\vec{\tau}) 
&= \delta\bigg(\!y_m \!-\!\sum_{k=1}^K \alpha_k \exp\Big(\jj g_m z_{mk} \! -\! \frac{g_m^2 \tau_k}{2}\Big)\!\bigg)
\label{eq:pyz},
\end{align}
where 
$\vec{\tau}\defn[\tau_1,\dots,\tau_K]\tran$ and $\vec{\alpha}\defn[\alpha_1,\dots,\alpha_K]\tran$ 
are hyperparameters of the GLM that will be estimated from $\vec{y}$.
%and $p(\vec{y}|\vec{Z}) = \prod_{m=1}^M \pyz(y_m|\vec{z}_m)$.

For the CL-AMP framework, any prior of the form 
\begin{align}
p(\vec{C}) = \prod_{n=1}^N \pc(\vec{c}_n\tran)
\label{eq:pc}
\end{align}
is admissible, where (with some abuse of notation) $\vec{c}_n\tran$ denotes the $n$th row of $\vec{C}$.
For all experiments in \secref{num}, we used the trivial prior $p(\vec{C})\propto 1$.

In summary, CL-AMP aims to compute the MMSE estimate of $\vec{C}\in\Real^{N\times K}$ from the sketch $\vec{y}\in\Complex^M$ under 
the prior $\vec{C}\sim \prod_{n=1}^N \pc(\vec{c}_n)$ from \eqref{pc} and 
the likelihood $\vec{y}\sim \prod_{m=1}^M \pyz(y_m|\vec{z}_m;\vec{\alpha},\vec{\tau})$ from \eqref{pyz}, where $\vec{z}_m\tran$ is the $m$th row of $\vec{Z}=\vec{AC}\in\Real^{M\times K}$ and $\vec{A}\in\Real^{M\times N}$ is a large random matrix with rows $\{\vec{a}_m\tran\}$ distributed uniformly on the unit sphere.
CL-AMP estimates the values of $\vec{\alpha}$ and $\vec{\tau}$ from the sketch prior to estimating $\vec{C}$, as detailed in the sequel.

\blue
%\subsection{Frequency Generation} \label{sec:freq_gen}

As proposed in \cite{Keriven:II:17}, 
%$\vec{a}_m$ were drawn uniformly on the unit sphere and 
the row-norms $\{g_m\}$ from \eqref{am} were drawn i.i.d.\ from the distribution 
\begin{equation}
p(g;\sigma^2) \propto 1_{[0,\infty)}(g) \sqrt{g^2\sigma^2 + \frac{g^4\sigma^4}{4}} \exp\Big(-\frac{1}{2}g^2\sigma^2\Big)
\label{eq:pg} 
\end{equation}
with shape parameter $\sigma^2$.
The authors in \cite{Keriven:II:17} suggest using $\sigma^2 = \frac{1}{NK}\sum_{k=1}^K \tr(\vec{\Phi}_k)$ and propose a method to estimate $\sigma^2$ from $\vec{y}$.
However, our numerical experiments suggest that the simpler assignment 
%\begin{equation}
%\sigma^2 
%= \frac{1}{N}\E\{\norm{\vec{x}}_2^2\} 
%\approx \frac{\norm{\vec{X}}_{F}^2}{NT}
%\label{eq:mom2}
%\end{equation}
\begin{equation}
\sigma^2 
= \E\{\norm{\vec{x}}_2^2\}/N 
\approx \norm{\vec{X}}_{F}^2/NT
\label{eq:mom2}
\end{equation}
provides significantly improved performance.
Note that the right side of \eqref{mom2} can be computed in an online manner, or approximated using a subset of the data.

\black

\subsection{Approximate Message Passing} \label{sec:amp}

Exactly computing the MMSE estimate of $\vec{C}$ from $\vec{y}$ is impractical due to the form of $\pyz$.
Instead, one might consider approximate inference via the sum-product algorithm (SPA), but even the SPA is intractable due to the form of $\pyz$.
Given the presence of a large random matrix $\vec{A}$ in the problem formulation, we instead leverage \emph{approximate message passing} (AMP) methods. %\cite{Donoho:PNAS:09,}.
In particular, we propose to apply the \emph{simplified hybrid generalized AMP} (SHyGAMP) methodology from \cite{Byrne:TSP:16}, while simultaneously estimating $\vec{\alpha}$ and $\vec{\tau}$ through expectation maximization (EM).
A brief background on AMP methods will now be provided to justify our approach.

The original AMP algorithm of Donoho, Maleki, and Montanari \cite{Donoho:PNAS:09} was designed to estimate i.i.d.\ $\vec{c}$ under the standard linear model (i.e., $\vec{y}=\vec{Ac}+\vec{n}$ with known $\vec{A}\in\Real^{M\times N}$ and additive white Gaussian noise $\vec{n}$).
The generalized AMP (GAMP) algorithm of Rangan \cite{Rangan:ISIT:11} extended AMP to the generalized linear model (i.e., $\vec{y}\sim p(\vec{y}|\vec{z})$ for $\vec{z}=\vec{Ac}$ and separable $p(\vec{y}|\vec{z})=\prod_{m=1}^M p(y_m|z_m)$).
Both AMP and GAMP give accurate approximations of the SPA under large i.i.d.\ sub-Gaussian $\vec{A}$, while maintaining a computational complexity of only $O(MN)$.
Furthermore, both can be rigorously analyzed via the state-evolution framework, which proves that they compute MMSE optimal estimates of $\vec{c}$ in certain regimes \cite{Bayati:TIT:11}. 

A limitation of AMP \cite{Donoho:PNAS:09} and GAMP \cite{Rangan:ISIT:11} is that they treat only problems with i.i.d.\ estimand $\vec{c}$ and separable likelihood $p(\vec{y}|\vec{z})=\prod_{m=1}^M p(y_m|z_m)$.
Thus, \emph{Hybrid GAMP} (HyGAMP) \cite{Rangan:TSP:17} was developed to tackle problems with a structured prior and/or likelihood.
HyGAMP could be applied to the compressive learning problem described in \secref{cl}, but it would require computing and inverting $O(N\!+\!M)$ covariance matrices of dimension $K$ at each iteration.
For this reason, we instead apply the \emph{simplified HyGAMP} (SHyGAMP) algorithm from \cite{Byrne:TSP:16}, which uses diagonal covariance matrices in HyGAMP to reduce its computational complexity. 
As described in \cite{Byrne:TSP:16}, SHyGAMP can be readily combined with the EM algorithm to learn the hyperparameters $\vec{\alpha}$ and $\vec{\tau}$.

%The original AMP algorithm and all of its derivatives have assumed the mixing matrix $\vec{W}$ is iid Gaussian.
%Our $\vec{W}$ does not fit this assumption, therefore we normalized of each $\vec{w}_m$ by its norm and absorbed this normalization into the likelihood $\pyz$ to more closely satisfy the iid Gaussian assumption.

\subsection{SHyGAMP} \label{sec:shygamp}

The SHyGAMP algorithm was proposed and described in detail in \cite{Byrne:TSP:16};
we provide only a brief review here.
\algref{shygamp} summarizes the SHyGAMP algorithm using the language of \secref{cl}. 
In lines~\ref{line:spaQc}-\ref{line:spachat}, with some abuse of notation, we use $\vec{c}_n\tran$ to denote the $n$th row of the centroid matrix $\vec{C}$ (where in \eqref{gm} we used $\vec{c}_k$ to denote the $k$th column of $\vec{C}$).
We also use
$\hvec{P}\defn[\hvec{p}_1,\dots,\hvec{p}_M]\tran$, 
$\hvec{Z}\defn[\hvec{z}_1,\dots,\hvec{z}_M]\tran$, 
$\hvec{R}\defn[\hvec{r}_1,\dots,\hvec{r}_N]\tran$,
$\oslash$ for componentwise division, and 
$\odot$ for componentwise multiplication.
In the sequel, covariance matrices will be denoted by (superscripted) $\vec{Q}$ and vectors of their diagonal elements denoted by (superscripted) $\vec{q}$.
A brief interpretation of SHyGAMP is now provided.

At each iteration, 
lines~\ref{line:spaQz}-\ref{line:spazhat} of \algref{shygamp} generate the posterior mean and covariance of the transform outputs $\vec{z}_m$ from \eqref{zm} under a likelihood $\pyz$ like \eqref{pyz} and the ``pseudo'' prior $\vec{z}_m\sim\mc{N}(\hvec{p}_m,\Qp)$, where $\hvec{p}_m$ and $\Qp=\Diag(\qp)$ are updated at each SHyGAMP iteration.
Thus, the pdf used for the covariance and expectation in lines~\ref{line:spaQz}-\ref{line:spazhat} is
\begin{align} 
\lefteqn{ 
\pzyp(\vec{z}_m|y_m,\hvec{p}_m;\Qp,\vec{\alpha},\vec{\tau}) 
}\nonumber\\
&= 
\frac{\pyz(y_m|\vec{z}_m;\vec{\alpha},\vec{\tau})\mc{N}(\vec{z}_m;\hvec{p}_m,\Qp)}
        {\int \pyz(y_m|\vec{z}'_m;\vec{\alpha},\vec{\tau})\mc{N}(\vec{z}'_m;\hvec{p}_m,\Qp) \dif \vec{z}'_m} .
\label{eq:pzyp}
\end{align}
Similarly, 
lines~\ref{line:spaQc}-\ref{line:spachat} compute the posterior mean and covariance of $\vec{c}_n$ under a prior $\pc$ of the form \eqref{pc} and ``pseudo'' measurements $\hvec{r}_n$ that follow the statistical model
\begin{align}
\hvec{r}_n=\vec{c}_n+\vec{v}_n, ~~ \vec{v}_n 
\sim\mc{N}(\vec{0},\Qr) ,
\end{align}
where $\hvec{r}_n$ and $\Qr = \Diag(\qr)$ are updated at each SHyGAMP iteration.
Thus, the pdf used for the covariance and expectation in lines~\ref{line:spaQc}-\ref{line:spachat} is 
\begin{align} 
\pcr(\vec{c}_n|\hvec{r}_n;\Qr) 
&= 
\frac{\pc(\vec{c}_n)\mc{N}(\vec{c}_n;\hvec{r}_n,\Qr)}
        {\int \pc(\vec{c}'_n)\mc{N}(\vec{c}'_n;\hvec{r}_n,\Qr) \dif \vec{c}'_n} 
\label{eq:pcr}.
\end{align} 

As the SHyGAMP iterations progress, the output $[\hvec{c}_1,\dots,\hvec{c}_N]\tran$ of \lineref{spachat} converges to an approximation of the MMSE estimate $\E\{\vec{C}|\vec{y}\}$, and the output $[\hvec{z}_1,\dots,\hvec{z}_M]\tran$ of \lineref{spazhat} converges to an approximation of the MMSE estimate $\E\{\vec{Z}|\vec{y}\}$.
Essentially, the SHyGAMP algorithm breaks an inference problem of dimension $NK$ into $O(M\!+\!N)$ inference problems of dimension $K$ \textb{(i.e., lines~\ref{line:spaQz}-\ref{line:spazhat} and \ref{line:spaQc}-\ref{line:spachat} of \algref{shygamp})}, each involving an independent-Gaussian pseudo-prior or pseudo-likelihood, evaluated iteratively.
The computational complexity of SHyGAMP is $O(MNK)$. 
%The full derivation of SHyGAMP can be found in \cite{Byrne:TSP:16}.

%Despite HyGAMP being a step in the right direction regarding computational feasibility, it still suffers from high complexity w.r.t.\ $K$. 
%In particular, every iteration of the HyGAMP algorithm requires the computation of $M+N$ $K$-dimensional covariance matrices and requires inverting $M$ of them.
%Therefore, we propose to use the Simplified-HyGAMP (SHyGAMP) algorithm, which simply approximates all covariance matrices as diagonal.
%This is easily implemented by initializing $\Qr_n(0)$ as a diagonal and just computing the diagonal of $\Qc_n$ and $\Qz_m$ in Lines \ref{line:spaQc} and \ref{line:spaQz} of \algref{shygamp}.
%The resulting algorithm's complexity (excluding the sketch) is $O\big(K(M+N)\big)$, which is linear in $M$, $N$, and $K$.

%The SHyGAMP is easily implemented by initializing $\Qr_n(0)$ as a diagonal matrix and just computing the diagonal of $\Qc_n$ and $\Qz_m$ in Lines \ref{line:spaQc} and \ref{line:spaQz} of \algref{shygamp}.

\begin{algorithm}[t]
\footnotesize
\caption{SHyGAMP}
\label{alg:shygamp}
\begin{algorithmic}[1]
\REQUIRE{
        Measurements $\vec{y}\in\Complex^M$,
        matrix $\vec{A}\in\Real^{M\times N}$ with $\|\vec{A}\|_F^2=M$, 
        pdfs $\pcr(\cdot|\cdot)$ and $\pzyp(\cdot|\cdot,\cdot;\vec{\alpha},\vec{\tau})$ from \eqref{pzyp} and \eqref{pcr},
        initial $\hvec{C}_0\in\Real^{N\times K}$ 
                and $\qp=\qp_0\in\Real_+^K$.}
        \vspace{1mm}
\STATE 
        $\hvec{S} \!\leftarrow\! \vec{0}$,
        $\hvec{C} \!\leftarrow\! \hvec{C}_0$.
\REPEAT    
        \vspace{0.5mm}
        \STATE 
        $\hvec{P} \leftarrow \vec{A} \hvec{C} - \hvec{S}\Diag(\qp)$ 
                        \label{line:phat}
        \STATE 
        $\qz_m \leftarrow \diag \big(\cov\big\{\vec{z}_m \biggiv y_m, \hvec{p}_m; \Diag(\qp),\vec{\alpha},\vec{\tau}\big\}\big),~m=1...M$ 
                        \label{line:spaQz}
        \STATE  
        $\hvec{z}_m \leftarrow \E\big\{\vec{z}_m \biggiv y_m, \hvec{p}_m; \Diag(\qp),\vec{\alpha},\vec{\tau}\big\},~m=1...M$ 
                        \label{line:spazhat}
        \STATE 
        $\qs \leftarrow \vec{1}\oslash\qp - \big(\frac{1}{M}\sum_{m=1}^M\qz_m\big) \oslash (\qp \odot \qp)$ 
                        \label{line:Qs}
        \STATE 
        $\hvec{S} \leftarrow (\hvec{Z} - \hvec{P} ) \Diag(\qp)^{-1} 
        $ 
                        \label{line:shat}
        \STATE 
        $\qr \leftarrow \frac{N}{M}\vec{1} \oslash \qs$
                        \label{line:Qr}
        \STATE 
        %$\hvec{R} \leftarrow \hvec{C} + \vec{A} \hvec{S}\tran \Diag(\qr)$ 
        $\hvec{R} \leftarrow \hvec{C} + \vec{A}\tran \hvec{S} \Diag(\qr)$ 
                        \label{line:rhat}
        \STATE 
        $\qc_n \leftarrow \diag \big(\cov\big\{\vec{c}_n \biggiv \hvec{r}_n; \Diag(\qr)\big\}\big),~n=1...N$ 
                        \label{line:spaQc}
        \STATE  
        $\hvec{c}_n \leftarrow \E\big\{\vec{c}_n \biggiv \hvec{r}_n; \Diag(\qr)\big\},~n=1...N$ 
                        \label{line:spachat}
        \STATE 
        $\qp \leftarrow \frac{1}{N}\sum_{n=1}^N \qc_n$ 
                        \label{line:Qp}
        \vspace{1mm}
\UNTIL{convergence}
\RETURN{$\hvec{C}$}
\end{algorithmic}
\end{algorithm}

\subsection{From SHyGAMP to CL-AMP}

The SHyGAMP algorithm can be applied to many different problems via appropriate choice of $\pyz$ and $\pc$.
To apply SHyGAMP to sketched clustering, we choose $\pyz$ and $\pc$ as described in \secref{cl}.
As we will see, the main challenge is evaluating lines \ref{line:spaQz}-\ref{line:spazhat} of \algref{shygamp} for the $\pyz$ in \eqref{pyz}.

\subsubsection{Inference of $\vec{z}_m$}

For lines~\ref{line:spaQz}-\ref{line:spazhat} of \algref{shygamp}, we would like to compute the mean and variance
\begin{align}
\hat{z}_{mk}
&= \frac{\int_{\Real^K} z_{mk} \pyz(y_m|\vec{z}_m) \mc{N}\big(\vec{z}_m;\hvec{p}_m,\Qp\big) \dif\vec{z}_m}{C_m} \\
q^{\rvz}_{mk}
&= \frac{ \int_{\Real^K} (z_{mk}\!-\!\hat{z}_{mk})^2 \pyz(y_m|\vec{z}_m) \mc{N}\big(\vec{z}_m;\hvec{p}_m,\Qp\big) \dif\vec{z}_m 
}{C_m} ,
\end{align}
where $q^{\rvz}_{mk}$ is the $k$th element of $\vec{q}_m^{\rvz}$ and
\begin{equation}
C_m 
= \int_{\Real^K} \pyz(y_m|\vec{z}_m) \mc{N}\big(\vec{z}_m;\hvec{p}_m,\Qp\big) \dif\vec{z}_m.
\end{equation}
However, due to the form of $\pyz$ in \eqref{pyz}, we are not able to find closed-form expressions for $\hat{z}_{mk}$ or $q^{\rvz}_{mk}$.
\textb{
Thus, we propose to approximate $\hat{z}_{mk}$ and $q^{\rvz}_{mk}$ 
by writing \eqref{ym3} as
\begin{align}
y_m 
&= \alpha_k \exp(-g_m^2 \tau_k / 2)\exp\big(\jj g_m z_{mk}\big) \nonumber \\
&\quad + \sum_{l\neq k} \alpha_l \exp(-g_m^2 \tau_l / 2) \exp\big(\jj g_m (z_{ml})\big)
\label{eq:gaussian} 
\end{align}
and treating the sum over $l$ as complex Gaussian.
}
For the remainder of this section, we suppress the subscripts ``$m$'' and ``$\ry|\rvecz$'' to simplify the notation.

\textb{We now give a brief sketch of the derivation.  First, we write}
%We begin by writing 
\eqref{gaussian} as
\begin{align}
y 
&= \underbrace{\alpha_k \exp(-g^2 \tau_k / 2)}_{\displaystyle \defn \beta_k} \exp\big(\jj\underbrace{g (z_k + n_k)}_{\displaystyle \defn \theta_k}\big) \nonumber \\
&\quad + \sum_{l\neq k} \underbrace{ \alpha_l \exp(-g^2 \tau_l / 2) }_{\displaystyle =\beta_l}  \underbrace{ \exp\big(\jj g (z_l+n_l)\big)}_{\displaystyle \defn v_l}
\label{eq:gaussian0} .
\end{align}
\textb{Here we introduce i.i.d.\ $n_k\sim\mc{N}(0,\qns)$, 
which will allow us to leverage the Gaussian multiplication rule (see footnote~\ref{foot:GMR}) to bypass tedious linear algebra.
Eventually we will take $\qns\rightarrow 0$, so that \eqref{gaussian0} matches \eqref{gaussian}.
Next we derive expressions \eqref{zhat} and \eqref{qz}, which state $\hat{z}_{k}$ and $q^{\rvz}_{k}$ in terms of the posterior mean and variance on the $2\pi$-periodic quantity $\theta_k$ in \eqref{gaussian0}.
By approximating the second term in \eqref{gaussian0} as Gaussian, the posterior of $\theta_k$ takes the form of a generalized von Mises distribution, as summarized in \eqref{gvm_post}.
Because the posterior mean and variance of $\theta_k$ are not computable in closed-form, we approximate them using numerical integration.
Finally, we relate the posterior mean and variance of $\theta_k$ back to $\hat{z}_k$ and $q^{\rvz}_k$.
}

\textb{We now begin the derivation. First,}
%Next 
we derive an expression for the marginal posterior $p(z_k|y)$ under the pseudo-prior $z_k\sim \mc{N}(\hat{p}_k,\qps_k)~\forall k$.
To start,
\begin{align}
p(z_k|y)
&= \int_{\Real^K} p(\vec{z},\theta_k|y) \dif\theta_k \dif\vec{z}\notk \\
&= \frac{1}{p(y)} \int_{\Real^K} p(y|\vec{z},\theta_k) p(\theta_k|\vec{z}) p(\vec{z}) 
        \dif\theta_k \dif\vec{z}\notk \\
&= \frac{1}{p(y)} \int_{\Real^K} p(y|\vec{z}\notk,\theta_k) \mc{N}(\theta_k;gz_k,g^2 \qns) \nonumber \\
&\qquad \times
\prod_{l=1}^K \mc{N}(z_l;\hat{p}_l,\qps_l) 
        \dif\theta_k \dif\vec{z}\notk ,
\end{align}
where $\vec{z}\notk\defn [z_1,\dots,z_{k-1},z_{k+1},\dots,z_K]\tran$.
A change-of-variables from $z_l$ to $\tilde{z}_l\defn z_l-\hat{p}_l$ for all $l\neq k$ gives
\begin{align}
p(z_k|y)
&= \frac{\mc{N}(z_k;\hat{p}_k,\qps_k)}{p(y)} 
        \int_\Real  \mc{N}(\theta_k;gz_k,g^2 \qns)  
\label{eq:post1}\\&\quad\times
        \left[ \int_{\Real^{K-1}} p(y|\tvec{z}\notk,\theta_k) 
        \prod_{l\neq k} \mc{N}(\tilde{z}_l;0,\qps_l) 
        \dif\tvec{z}\notk \right]
        \dif\theta_k 
\nonumber ,
\end{align}
where $p(y|\tvec{z}\notk,\theta_k)$ is associated with the generative model
\begin{align}
y &= \scalek \exp(\jj\theta_k) + \sum_{l\neq k} 
       \scalel \exp\big(\jj g (\hat{p}_{l} + \tilde{z}_l + n_l )\big)
\label{eq:gaussian1}
\end{align}
with i.i.d.\ $n_l \sim \mc{N}(0,\qns)$.
%Note that $p(y|\tvec{z}\notk,\theta_k)$ is $2\pi$-periodic in $\theta_k$ and $g\tilde{z}_l$ for all $l\neq k$.
Now, because $\tilde{z}_l$ and $n_l$ are (apriori) mutually independent zero-mean Gaussian variables, we can work directly with the sum $\tilde{n}_l\defn\tilde{z}_l+n_l\sim\mc{N}(0,\qps_l+\qns)$ and thus bypass the inner integral in \eqref{post1}.
This allows us to write
\begin{align}
p(z_k|y)
&= \frac{\mc{N}(z_k;\hat{p}_k,\qps_k)}{p(y)} 
        \int_\Real \mc{N}(\theta_k; g z_k, g^2 \qns)
        p(y|\theta_k) 
        \dif\theta_k ,
\label{eq:post2}
\end{align} 
where $p(y|\theta_k)$ is associated with the generative model
\begin{align}
y 
&= \scalek \exp(\jj\theta_k)  
        + \sum_{l\neq k} \scalel \underbrace{\exp(\jj g (\hat{p}_{l} + \tilde{n}_l))}_{\displaystyle = v_l}
\label{eq:gaussian2}
\end{align}
with i.i.d.\ $\tilde{n}_l\sim\mc{N}(0,\qps_l+\qns)$.
%Note that $p(y|\theta_k)$ is $2\pi$-periodic in $\theta_k$.
Recalling that $y\in\Complex$, it will sometimes be useful to write \eqref{gaussian2} as
\begin{align}
\mat{\real\{y\}\\
     \imag\{y\}} 
&\sim \mc{N}\bigg(
        \scalek \mat{\cos(\theta_k)\\
             \sin(\theta_k)}
        + \sum_{l\neq k} 
        \scalel \E\bigg\{\mat{\real\{v_l\}\\
                      \imag\{v_l\}} \bigg\},~  \nonumber\\&\quad\quad\quad
        \sum_{l\neq k} 
        \scalell \cov\bigg\{\mat{\real\{v_l\}\\
                      \imag\{v_l\}} \bigg\} \bigg) 
\label{eq:gaussian3}.
\end{align}

To compute the posterior mean of $z_k$, \eqref{post2} implies 
\begin{align}
\lefteqn{
\hat{z}_k \defn \E\{z_k|y\} = \int_\Real z_k \,p(z_k|y) \dif z_k }\\
&= \frac{1}{p(y)} \int_\Real \left[\int_\Real z_k \,\mc{N}({g}z_k;\theta_k ,{g^2}\qns) 
\mc{N}(z_k;\hat{p}_k,\qps_k) \dif z_k \right] \nonumber \\
& \quad \times                         
   p(y|\theta_k) \dif \theta_k \\
&= \int_\Real
    \left[\int_\Real z_k 
    \,\mc{N}\!\left(z_k;
    \frac{\frac{\theta_k {/g}}{\qns}+\frac{\hat{p}_k}{\qps_k}}
         {\frac{1}{\qns}+\frac{1}{\qps_k}},
    \frac{1}{\frac{1}{\qns}+\frac{1}{\qps_k}}
    \right)
    \dif z_k \right] \nonumber \\
    &\quad \times
    \underbrace{ 
   \frac{\mc{N}\Big(\theta_k ;{g \hat{p}_k},{g^2 (\qns+\qps_k)}\Big) \,p(y|\theta_k)}{p(y)}
   }_{\displaystyle =p(\theta_k|y)}
   \dif \theta_k \label{eq:postmean1}\\
&= \int_\Real
    \frac{\frac{\theta_k {/ g}}{\qns}+\frac{\hat{p}_k}{\qps_k}}
         {\frac{1}{\qns}+\frac{1}{\qps_k}}\,
   p(\theta_k|y) \dif \theta_k \\
&= \frac{\hat{p}_k}{\qps_k/\qns+1}
   + \frac{\hat{\theta}_k {/ g}}{1+\qns/\qps_k} 
   \text{~~for~}
   \hat{\theta}_k \defn
   \int_\Real \theta_k \,p(\theta_k|y) \dif \theta_k 
\label{eq:zhat} ,
\end{align}
where the Gaussian pdf multiplication rule\footnote{\textb{According to the Gaussian multiplication rule, we have} $\mc{N}(\vec{x};\vec{a},\vec{A})\mc{N}(\vec{x};\vec{b},\vec{B}) = \mc{N}(\vec{0};\vec{a}-\vec{b}, \vec{A}+\vec{B})\mc{N}\big(\vec{x};(\vec{A}^{-1} + \vec{B}^{-1})^{-1}(\vec{A}^{-1}\vec{a} + \vec{B}^{-1}\vec{b}), (\vec{A}^{-1} + \vec{B}^{-1})^{-1}\big)$.\label{foot:GMR}} was used in \eqref{postmean1} and where $\hat{\theta}_k$ denotes the posterior mean of $\theta_k$.

For the posterior variance of $z_k$, a similar approach gives
\begin{align}
\qzs_k
&\defn \var\{z_k|y\} 
=\int_\Real \big(z_k-\hat{z}_k\big)^2 \, p(z_k|y) \dif z_k \\
&= \frac{1}{p(y)} \int_\Real
        \bigg[\int_\Real (z_k\!-\!\hat{z}_k)^2 \,\mc{N}({g}z_k;\theta_k,{g^2}\qns) \nonumber \\
                         &\quad \times \mc{N}(z_k;\hat{p}_k,\qps_k) \dif z_k \bigg] p(y|\theta_k) \dif \theta_k \\
&= \int_\Real
    \left[\int_\Real (z_k\!-\!\hat{z}_k)^2
    \,\mc{N}\!\left(z_k;
    \frac{\frac{\theta_k {/ g}}{\qns}\!+\!\frac{\hat{p}_k}{\qps_k}}
         {\frac{1}{\qns}\!+\!\frac{1}{\qps_k}},
    \frac{1}{\frac{1}{\qns}\!+\!\frac{1}{\qps_k}}
    \right)
    \dif z_k \right] \nonumber \\
    &\quad\times
   p(\theta_k|y) \dif \theta_k .
\end{align}
Using a change-of-variables from $z_k$ to $\tilde{z}_k\defn z_k-\hat{z}_k$, we get
\begin{align}
\qzs_k
&= \int_\Real
    \left[\int_\Real \tilde{z}_k^2
    \,\mc{N}\!\left(\tilde{z}_k;
    \frac{\frac{\theta_k {/ g}}{\qns}\!-\!\frac{\hat{\theta}_k {/ g}}{\qns}}
         {\frac{1}{\qns}\!+\!\frac{1}{\qps_k}},
    \frac{1}{\frac{1}{\qns}\!+\!\frac{1}{\qps_k}}
    \right)
    \dif \tilde{z}_k \right] \nonumber \\
    &\quad \times
   p(\theta_k|y) \dif \theta_k \\
&= \int_\Real \left[
    \left(\frac{(\theta_k-\hat{\theta}_k){/ g}}{1+\qns/\qps_k}\right)^2
    + \frac{\qns}{1+\qns/\qps_k}
    \right]
   p(\theta_k|y) \dif \theta_k \\
&= \frac{\qns}{1\!+\!\qns/\qps_k} \!+\!
   {\frac{1}{g^2}}\left(\frac{1}{1\!+\!\qns/\qps_k}\right)^2
   \underbrace{
   \int_\Real (\theta_k\!-\!\hat{\theta}_k)^2 \,
   p(\theta_k|y) \dif \theta_k 
   }_{\displaystyle \defn q^{\theta}_k = \var\{\theta_k|y\}}  
\label{eq:qz} .
\end{align}

The computation of $\hat{z}_k$ and $\qzs_k$ is still complicated by the form of the posterior $p(\theta_k|y)$ implied by \eqref{gaussian2}.
To circumvent this problem, we propose to apply a Gaussian approximation to the sum in \eqref{gaussian2}.
%The Gaussian approximation of $p(y|\theta_k)$ is fully parameterized by the mean and covariance of the real and imaginary components of $y|\theta_k$.
Because $\{\tilde{n}_l\}_{\forall l\neq k}$ are mutually independent, the mean and covariance of the sum in \eqref{gaussian2} are simply the sum of the means and covariances (respectively) of the $K-1$ terms making up the sum.
Recalling \eqref{gaussian3}, this implies that 
\begin{align}
p\bigg(\mat{\real\{y\}\\\imag\{y\}}\bigg|\theta_k\bigg)
&\approx \mc{N}\bigg(\mat{\real\{y\}\\\imag\{y\}};
         \scalek \mat{\cos(\theta_k)\\\sin(\theta_k)} + 
         \vec{\mu}_k,\vec{\Sigma}_k\bigg)
\label{eq:likelihood_theta1}
\end{align}
with 
\begin{align}
{\blue\vec{\mu}_k}
&
{\blue
= \sum_{l\neq k} {\alpha_l} e^{-g^2(\tau_l + \qps_l)/2} \mat{\cos(g\hat{p}_l)\\\sin(g\hat{p}_l)} }
\label{eq:mu_k_l1}  \\
\vec{\Sigma}_k
&= \frac{1}{2}\sum_{l\neq k}\scalell \big(1-e^{-g^2 \qps_l} \big) \nonumber \\
&\quad\times
   \bigg( \vec{I} - e^{-g^2 \qps_l} \mat{
   \cos(2g\hat{p}_l) & \sin(2g\hat{p}_l) \\
   \sin(2g\hat{p}_l) &-\cos(2g\hat{p}_l) }
   \bigg) .
    \label{eq:Sigma_k_l4} 
\end{align}
We note that \eqref{mu_k_l1} and \eqref{Sigma_k_l4} were obtained using
\begin{align}
\E\big\{\real\{v_l\}\big\} 
&= \exp\big(-g^2\qps_l/2 \big) \cos(g\hat{p}_l) \\
\E\big\{\imag\{v_l\}\big\} 
&= \exp\big(-g^2\qps_l/2 \big) \sin(g\hat{p}_l) \\
2\E\big\{\real\{v_l\}^2\big\} 
&= 1 + \exp\big(-g^2\qps_l\big)\cos(2g\hat{p}_l) \\
2\E\big\{\imag\{v_l\}^2\big\} 
&= 1 - \exp\big(-g^2\qps_l\big)\cos(2g\hat{p}_l) \\  
2\E\big\{\real\{v_l\}\imag\{v_l\}\big\} 
&= \exp\big(-g^2\qps_l\big)\sin(2g\hat{p}_l),
\end{align}
which use the fact that, after letting $\qns\rightarrow 0$,  
\begin{align}
\E \{v_l\} &= \int_{\Real} \mc{N}(z_l;\hat{p}_l, \qps_l) \exp(\jj g z_l) \dif z_l \\
&= \exp\big(\jj g \hat{p}_l - g^2 \qps_l/2\big).
\label{eq:exp_v}
\end{align}

Rewriting \eqref{likelihood_theta1} as
\begin{align}
p\bigg(&\beta_k^{-1}\mat{\real\{y\}\\\imag\{y\}}\bigg|\theta_k\bigg) 
\label{eq:likelihood_theta2} \\
&\approx \mc{N}\bigg(
         \mat{\cos(\theta_k)\\\sin(\theta_k)};
         \beta_k^{-1}\mat{\real\{y\}\\\imag\{y\}} -
         \beta_k^{-1}\vec{\mu}_k,\beta_k^{-2}\vec{\Sigma}_k\bigg) 
\nonumber ,
\end{align}
the right side of \eqref{likelihood_theta2} can be recognized as being proportional to the generalized von Mises (GvM) density over $\theta_k\in[0,2\pi)$ from \cite{Gatto:SM:07}.
Under this GvM approximation, we have \cite{Gatto:SM:07} that 
\begin{align}
p(y|\theta_k)
&\propto \exp\big(\kappa_k \cos(\theta_k-\zeta_k) + \bar{\kappa}_k\cos[2(\theta_k-\bar{\zeta}_k)] \big) 
\label{eq:pytheta}
\end{align}
for parameters $\kappa_k,\bar{\kappa}_k>0$ and $\zeta_k,\bar{\zeta}_k\in[0,2\pi)$ defined from $\beta_k^{-1}y$, $\beta_k^{-1}\vec{\mu}_k$, and $\beta_k^{-2}\vec{\Sigma}_k$.
In particular,
\begin{align}
\kappa_k \cos(\zeta_k)
&= -\frac{1}{1-\rho_k^2}\bigg(\frac{\rho_k \bar{\nu}_k}{\sigma_k\bar{\sigma}_k}-\frac{\nu_k}{\sigma_k^2}\bigg) \\
\kappa_k \sin(\zeta_k)
&= -\frac{1}{1-\rho_k^2}\bigg(\frac{\rho_k \nu_k}{\sigma_k\bar{\sigma}_k}-\frac{\bar{\nu}_k}{\bar{\sigma}_k^2}\bigg) \\
\bar{\kappa}_k \cos(2\bar{\zeta}_k)
&= -\frac{1}{4(1-\rho_k^2)}\bigg(\frac{1}{\sigma_k^2}-\frac{1}{\bar{\sigma}_k^2}\bigg)\\
\bar{\kappa}_k \sin(2\bar{\zeta}_k)
&= \frac{\rho_k}{2(1-\rho_k^2)\sigma_k\bar{\sigma}_k} ,
\end{align}
where
\begin{align}
\mat{\nu_k\\\bar{\nu}_k}
&\defn \beta_k^{-1} \bigg( \mat{\real\{y\}\\\imag\{y\}} - \vec{\mu}_k \bigg) \\
\mat{\sigma_k^2 & \rho_k \sigma_k\bar{\sigma}_k\\
     \rho_k \sigma_k\bar{\sigma}_k & \bar{\sigma}_k^2}
&\defn \beta_k^{-2} \vec{\Sigma}_k. 
\end{align}
From \eqref{pytheta} and the SHyGAMP pseudo-prior $z_k\sim\mc{N}(\hat{p}_k,\qps_k)$, we see that the posterior on $\theta_{k}$ takes the form 
\begin{align}
\lefteqn{ p(\theta_{k}|y)
\propto \mc{N}\big(\theta_{k} ;g \hat{p}_{k} , g^2 \qps_k \big) \,p(y|\theta_{k}) }
\\
&\propto \exp\!\bigg[ 
\kappa_k \cos(\theta_{k}-\zeta_k) + \bar{\kappa}_k\cos[2(\theta_{k}-\bar{\zeta}_k)] 
- \frac{(\theta_{k}-{g\hat{p}_{k} })^2}{2g^2\qps_k} 
\bigg] 
\label{eq:gvm_post} 
.
\end{align}

We now face the task of computing $\hat{\theta}_k = \E\{\theta_{k}|y\}$ and $q^{\theta}_k = \var\{\theta_{k}|y\}$ under \eqref{gvm_post}.
Since these quantities do not appear to be computable in closed form, we settle for an approximation, such as that based on the Laplace approximation \cite{Bishop:Book:07} or numerical integration.
For the Laplace approximation, we would first compute $\hat{\theta}\MAPk \defn \argmax_{\theta_k} \ln p(\theta_{k}|y)$ 
and then approximate 
$\hat{\theta}_k \approx \hat{\theta}\MAPk$ and 
$q^{\theta}_k \approx - \frac{\dif^2}{{\dif \theta_k}^2} \ln p(\theta_{k}|y)\big|_{\theta_k = \hat{\theta}\MAPk} $.
However, since computing $\argmax_{\theta_k} \ln p(\theta_{k}|y)$ is complicated due to the presence of multiple local maxima, we instead use numerical integration.
For this, we suggest a grid of $N_{\text{pts}} N_{\text{per}} + 1$ uniformly-spaced points centered at $g\hat{p}_k$ with width $2\pi N_{\text{per}}$, where $N_{\text{per}}=\Big\lceil \frac{N_{\text{std}}}{\pi}\sqrt{g^2 \qps_k}\Big\rceil$.
This choice of grid ensures that the sampling points cover at least $N_{\text{std}}$ standard deviations of the prior on $\theta_k$.
We used $N_{\text{std}}=4$ and $N_{\text{pts}} = 7$ in the numerical experiments in \Secref{num}.
%Finally, we compute $\E\{\theta_{k}|y\} \approx (\theta_{k2} - \theta_{k1})\sum_{n=1}^{N_{\text{points}} N_{\text{periods}} + 1} \theta_{kn} p(\theta_{kn}|y)$ and $\var\{\theta_k|y\} = (\theta_{k2} - \theta_{k1})\sum_{n=1}^{N_{\text{points}} N_{\text{periods}} + 1} \theta_{kn}^2 p(\theta_{kn}|y) - \E\{\theta_{k}|y\}^2$, after which we $\hat{z}_k = \E\{\theta_{k}|y\} / g$ and $q^{\rvz}_k = \var\{\theta_{k}|y\} / g^2$.

Finally, after approximating $\hat{\theta}_k$ and $q^{\theta}_k$ via numerical integration, we set $\hat{z}_k =\hat{\theta}_k / g$ and $q^{\rvz}_k = q^{\theta}_k / g^2$.

\subsubsection{Inference of $\vec{c}_n$}

Recall that lines~\ref{line:spaQc}-\ref{line:spachat} of \algref{shygamp} support an arbitrary prior $\pc$ on $\vec{c}_n$.
For the experiments in \secref{num}, we used the trivial non-informative prior $\pc(\vec{c}_n)\propto 1$, after which lines~\ref{line:spaQc}-\ref{line:spachat} reduce to
\begin{align}
\qc_n &= \qr~\forall n
\text{~~and~~}
\hvec{c}_n = \hvec{r}_n ~\forall n .
\end{align}
%Thus $\pcr$ is Gaussian and the posterior mean and covariance of $\rvecc_n$ can be computed straightforwardly as 
%\begin{align}
%\Qc_n &= \big(\nu^{-1}\vec{I} + [\Qr]^{-1}\big)^{-1} \defn \Qc\\
%\hvec{c}_n &= \Qc [\Qr]^{-1}\hvec{r}_n
%\end{align}
%Above, \textb{the inversions} are simplified by the fact that $\Qr$ is diagonal.
%For the experiments in \secref{num}, we let $\nu \rightarrow \infty$ \textb{so that the prior is non-informative}, in which case $\Qc = \Qr$ and $\hvec{c}_n=\hvec{r}_n$.

\subsection{Initialization} \label{sec:init}

We recommend initializing CL-AMP with $\hvec{C}=\hvec{C}_0$ and $\qp=\qp_0$, where $\hvec{C}_0$ is drawn i.i.d.\ $\mc{N}(0,\sigma^2)$ and where $\qp_0=\sigma^2\vec{1}$, with $\sigma^2$ from \eqref{mom2} 
(as described in \secref{cl}).
%(as described in \secref{freq_gen}).

In some cases, running CL-AMP from $R>1$ different random initializations can \textb{help to avoid spurious} solutions.
Here, CL-AMP is run from a different random initialization $\hvec{C}_{0,r}$, for $r=1,\dots,R$, and then the quality of the recovered solution $\hvec{C}_r$ is evaluated by constructing the ``estimated sketch'' $\hvec{y}_r$ via 
\begin{equation}
\hat{y}_{mr} = \sum_{k=1}^K \alpha_{k} \exp(-g_m^2\tau_{k})\exp(\jj g_m \vec{a}_m\tran\hvec{c}_{kr})
\end{equation}
recalling \eqref{zm} and \eqref{ym3},
and then measuring its distance to the true sketch $\vec{y}$.
The initialization index is then selected as 
\begin{equation}
r_* = \argmin_r \norm{\vec{y}-\hvec{y}_r}, 
\end{equation}
and the centroids saved as $\hvec{C}=\hvec{C}_{r_*}$.
In \Secref{num}, we used $R=2$ for all experiments.

\subsection{Hyperparameter Tuning} \label{sec:tune}

The likelihood model $\pyz$ in \eqref{pyz} depends on the unknown hyperparameters $\vec{\alpha}$ and $\vec{\tau}$.
%Similarly, the prior $\pc$ depends on the unknown variance $\nu$.
We propose to estimate these hyperparameters using a combination of \emph{expectation maximization} (EM) and SHyGAMP, as suggested in \cite{Byrne:TSP:16} and detailed---for the simpler case of GAMP---in \cite{Vila:TSP:13}.
The idea is to run SHyGAMP using an estimate of $\vec{\alpha}$ and $\vec{\tau}$, update $\vec{\alpha}$ and $\vec{\tau}$ from the SHyGAMP outputs, and repeat until convergence.
For the first estimate, we suggest to use $\alpha_{k}=\frac{1}{K}$ and $\tau_{k}=0~\forall~k$.

Extrapolating \cite[eq. (23)]{Vila:TSP:13} to the SHyGAMP case, the EM update of $(\vec{\alpha},\vec{\tau})$ takes the form
\begin{align}
(\hvec{\alpha}, \hvec{\tau}) 
&= \hspace{-3mm}\argmax_{\vec{\alpha} \geq \vec{0}, \vec{\alpha}\tran\vec{1} = 1, \vec{\tau}>\vec{0}}  
\sum_{m=1}^M \int_{\Real^K} \mc{N}(\vec{z}_m; \hvec{z}_m, \Qz_m) 
\label{eq:EM_0} \\ &\hspace{30mm} \times
\ln \pyz(y_m | \vec{z}_m;\vec{\alpha},\vec{\tau}) 
\dif \vec{z}_m
\nonumber ,
\end{align}
where $\hvec{z}_m$ and $\Qz_m = \Diag\{\qz_m\}$ are obtained by running SHyGAMP to convergence under $(\vec{\alpha},\vec{\tau})$.
To proceed, we model the Dirac delta in \eqref{pyz} 
%for $\pyz(y_m|\vec{z}_m; \vec{\alpha},\vec{\tau})$ 
using a circular Gaussian pdf with vanishingly small variance $\epsilon>0$, in which case
\begin{align}
\lefteqn{ \ln \pyz(y_m|\vec{z}_m;\vec{\alpha},\vec{\tau}) }
\label{eq:quad_approx}\\
&= -\frac{1}{\epsilon}\bigg|y_m - \sum_{k=1}^K \alpha_k \exp\bigg(\jj g_mz_{mk} - \frac{g_m^2\tau_k}{2}\bigg)\bigg|^2 + \text{const} 
\nonumber .
\end{align}
Plugging \eqref{quad_approx} back into \eqref{EM_0}, we see that the constant and the $1/\epsilon$-scaling play no role in the optimization, and so we can discard them to obtain
\begin{align}
(\hvec{\alpha}, \hvec{\tau}) 
&= \hspace{-3mm}\argmin_{\vec{\alpha} \geq \vec{0}, \vec{\alpha}\tran\vec{1} = 1, \vec{\tau}>\vec{0}}  
\sum_{m=1}^M \int_{\Real^K} \mc{N}(\vec{z}_m; \hvec{z}_m, \Qz_m) 
\label{eq:EM_1} \\ &\quad \times
\bigg|y_m - \sum_{k=1}^K \alpha_k \exp\bigg(\jj g_mz_{mk} - \frac{g_m^2\tau_k}{2}\bigg)\bigg|^2 
\dif \vec{z}_m
\nonumber .
\end{align}

A closed-form solution to the optimization problem in \eqref{EM_1} seems out of reach.
Also, the optimization objective is convex in $\vec{\alpha}$ for fixed $\vec{\tau}$, and convex in $\vec{\tau}$ for fixed $\vec{\alpha}$, but not jointly convex in $[\vec{\alpha}\tran,\vec{\tau}\tran]$.
Although the optimization problem \eqref{EM_1} is difficult to solve, the solutions obtained by gradient projection (GP) \cite{Bertsekas:Book:99} seem to work well in practice.
Also, GP is made practical by closed-form gradient expressions.
In particular, let
\begin{align}
q_{mk} 
&\defn \exp\bigg(-\frac{g_m^2 \tau_k}{2}\bigg) \\
\rho_{mk} 
&\defn \exp\bigg(\jj g_m \hat{z}_{mk}\! -\! \frac{\qzs_{mk} g_m^2}{2}\bigg) ,
\end{align} 
and recall that $v_{mk}=\exp(\jj g_m z_{mk})$ from \eqref{gaussian0}
\textb{(although there the $m$ subscript was suppressed)}.
%Note that $q_{mk}$ depends on $\tau_k$.
Then the $m$th term of the sum in the objective in \eqref{EM_1} becomes
\begin{align}
&\int_{\Real^K} \mc{N}(\vec{z}_{m}; \hvec{z}_{m},\Qz_m) \bigg|y_m - \sum_{k=1}^K \alpha_k q_{mk} v_{mk}\bigg|^2 \dif \vec{z}_m   \nonumber \\[-2mm]
&= |y_m|^2 - 2  \sum_{k=1}^K \alpha_k q_{mk} \real\big\{y_m^* \rho_{mk} \big\} \nonumber \\[-2mm]
&\phantom{==} + \sum_{k=1}^K \alpha_{k} q_{mk} \rho^*_{mk} \sum_{l \neq k}^K  \alpha_{l}  q_{ml} \rho_{ml}  + \sum_{k=1}^K \alpha_{k}^2 q_{mk}^2,
\end{align}
where we used the fact that $\int_\Real \mc{N}(z_{mk};\hat{z}_{mk}, \qzs_{mk})v_{mk}\dif z_{mk} = \rho_{mk}$.
After reapplying the sum over $m$, we get
\begin{align}
&\frac{\partial}{\partial \alpha_k} \sum_{m=1}^M \int_{\Real^K} \mc{N}(\vec{z}_{m}; \hvec{z}_{m},\Qz_m)\bigg|y_m\!-\!\sum_{k=1}^K \alpha_k q_{mk} v_{mk}\bigg|^2 \dif \vec{z}_m \nonumber \\ 
&= -2 \sum_{m=1}^M q_{mk} \gamma_{mk}
\label{eq:EM_4} \\
&\frac{\partial}{\partial \tau_k} \sum_{m=1}^M \int_{\Real^K} \mc{N}(\vec{z}_{m}; \hvec{z}_{m}, \Qz_m) \bigg|y_m\!-\!\sum_{k=1}^K \alpha_k q_{mk} v_{mk}\bigg|^2 \dif \vec{z}_m \nonumber \\
&= \alpha_k \sum_{m=1}^M g_m^2 q_{mk} \gamma_{mk}
\label{eq:EM_5} 
\end{align}
for 
\begin{align}
\gamma_{mk}
&\defn \real\big\{y_m^* \rho_{mk}\big\} - \alpha_k q_{mk} - \sum_{l\neq k}^K \alpha_{l} q_{ml} \real \big\{\rho^*_{mk} \rho_{ml}\big\} .
\end{align}

We found that complexity of hyperparameter tuning can be substantially reduced, without much loss in accuracy, by using only a subset of the terms in the sum in \eqref{EM_1}, as well as in the corresponding gradient expressions \eqref{EM_4}-\eqref{EM_5}.
For the experiments in \secref{num}, we used a fixed random subset of $\min(M,20K)$ terms.
%\textb{We chose this approach over minibatch since it converged more quickly than minibatch.}
%\textb{[Use minibatch?]}

%In practice this approach typically converges in fewer than 10 ``outer" iterations.

\subsection{Algorithm Summary}

\algref{tune} summarizes the CL-AMP algorithm with $R$ random initializations and tuning of the hyperparameters $(\vec{\alpha},\vec{\tau})$.
Note that the random initializations $\{\hvec{C}_{0,r}\}$ are used only for the first EM iteration, i.e., $i=0$.
Subsequent EM iterations (i.e., $i\geq 1$) are initialized using the output $\hvec{C}_i$ of the previous EM iteration.

\begin{algorithm}[t]
\footnotesize
\caption{CL-AMP with hyperparameter tuning and multiple random initializations}
\label{alg:tune}
\begin{algorithmic}[1]
\REQUIRE{     
        Measurements $\vec{y}\in\Complex^M$,
        gains $\{g_m\}_{m=1}^M$,
        %matrix $\vec{A}$ with $\|\vec{A}\|_F^2=M$, 
        %pdfs $\pcr$ and $\pzyp$ from \eqref{pcr}-\eqref{pzyp},
        number of initializations $R\geq 1$,
        initializations 
                $\{\hvec{C}_{0,r}\}_{r=1}^R$,
                $\qp_0$,
                $\vec{\alpha}_0$,
                $\vec{\tau}_0$.
        } \vspace{1mm}
\STATE $i = 0$
        \REPEAT
        \IF {$i=0$}
        \FOR {$r=1:R$}
		\STATE Run CL-AMP with fixed $(\vec{\alpha}_0$, $\vec{\tau}_0)$ from initialization $(\hvec{C}_{0,r},\qp_0)$, yielding output $\hvec{C}_{1,r}$, $\hvec{Z}_{r}$, and $\{\qz_{mr}\}_{m=1}^M$.
		\ENDFOR
		\STATE Compute $\hat{y}_{mr} \defn \sum_{k=1}^K \alpha_{0k} \exp(-g_m^2\tau_{0k})\exp(\jj g_m \hat{z}_{mkr})~\forall mr$%
		\STATE Find $r_* = \argmin_r \norm{\vec{y}-\hvec{y}_r}$.
		\STATE Set $\hvec{C}_{1} = \hvec{C}_{1,r_*}$, $\hvec{Z} = \hvec{Z}_{r_*}$ and  $\{\qz_m\}_{m=1}^M = \{\qz_{mr_*}\}_{m=1}^M$.
		\ELSE
		\STATE Run CL-AMP with fixed $(\vec{\alpha}_i,\vec{\tau}_i)$ from initialization $(\hvec{C}_i,\qp_0)$, yielding output $\hvec{C}_{i+1}$, $\hvec{Z}$, and $\{\qz_{m}\}_{m=1}^M$.
		\ENDIF		
		 \STATE Compute $(\vec{\alpha}_{i+1},\vec{\tau}_{i+1})$ via \eqref{EM_1} using $\hvec{Z}$ and $\{\qz_{m}\}_{m=1}^M$.
		 \label{line:em}
		\STATE $i \leftarrow i+1$.
\UNTIL{convergence}
\end{algorithmic}
\end{algorithm}

%-------------------------------------------------------------------------------
\section{Numerical Experiments} \label{sec:num}

In this section, we present the results of several experiments used to test the performance of the CL-AMP, CL-OMPR, and k-means++ algorithms.
For k-means++, we used the implementation provided by MATLAB
%with one replicate (i.e., 1 run from a random initialization)
and, for CL-OMPR, we downloaded the MATLAB implementation from \cite{clompr:toolbox}.
% and enabled the ``++" initialization method.
CL-OMPR and CL-AMP used the same sketch $\vec{y}$, whose frequency vectors $\vec{W}$ were drawn using the method described in 
\secref{cl}, 
%\secref{freq_gen}, 
with the scaling parameter $\sigma^2$ set via \eqref{mom2}.
For CL-OMPR and CL-AMP, the reported runtimes include the time of computing the sketch, unless otherwise noted.
All experiments were run on a Dell PowerEdge C6320 two-socket server with Intel Xeon E5-2680 v4 processors (14 cores, 2.40GHz) and 128GB RAM.

\subsection{Experiments with Synthetic Data} \label{sec:sse_min}

\subsubsection{Performance vs.\ sketch length $M$} \label{sec:sse_v_m}

In the first experiment, we test each algorithm's ability to minimize SSE on a set of training data, i.e., to solve the problem \eqref{sse}.
In addition, we test how well the recovered centroids work in minimum-distance classification.

The experiment was conducted as follows.
Fixing the number of classes at $K=10$ and the data dimension at $N=100$, 
ten Monte Carlo trials were performed. 
In each trial, the true centroids were randomly drawn\footnote{This data-generation model was chosen to match that from \cite{Keriven:II:17}, and is intended to have a relatively constant Bayes error rate w.r.t.\ $N$ and $K$.  \textb{For the chosen parameters, the Bayes error rate is extremely small: $10^{-24}$.  Thus, when the centroids are accurately recovered, the classification error rate should be essentially zero.}} 
as $\vec{c}_k\sim \mc{N}(\vec{0}_N, 1.5^2 K^{2/N} \vec{I}_N)$.
Then, using these centroids, a training dataset $\{\vec{x}_t\}_{t=1}^T$ with $T=10^7$ samples was drawn from the GMM \eqref{gm} with weights $\alpha_k = 1/K$ and covariances $\vec{\Phi}_k = \vec{I}_N \forall k$.
Additionally, a test dataset $\{\ovec{x}_t\}$ of $10^6$ samples was independently generated.

For centroid recovery, k-means++ was invoked on the training dataset, and both
CL-AMP and CL-OMPR were invoked after sketching the training data with $M$ samples as in \eqref{sk}.
Sketch lengths $M/KN\in\{1,2,3,5,10,20\}$ were investigated.
CL-AMP used two random initializations, i.e., $R=2$ as defined in \algref{tune}.

For each algorithm, the SSE of its estimated centroids $\{\hvec{c}_k\}_{k=1}^K$ was calculated using the training data $\{\vec{x}_t\}_{t=1}^T$ via \eqref{sse}.
Additionally, the performance of the estimated centroids in minimum-distance classification was evaluated as follows.
First, labels $\{j_k\}_{k=1}^K$ were assigned to the estimated centroids by solving the linear assignment problem \cite{Kuhn:NRLQ:55} without replacement, given by
\begin{equation}
\argmin_{\{j_1,...,j_K\}=\{1,...,K\}} \sum_{k=1}^K \norm{\vec{c}_k - \hvec{c}_{j_k}}^2_2
\label{eq:LAP}.
\end{equation}
Next, each test sample $\ovec{x}_t$ was classified using minimum-distance classification, producing the estimated label
\begin{equation}
\hat{k}_t = \argmin_{k\in\{1,...,K\}} \norm{\ovec{x}_t - \hvec{c}_{j_k}}.
\end{equation}
The classification error rate (CER) was then calculated as the proportion of estimated labels $\hat{k}_t$ that do not equal the true label $k_t$ from which the test sample $\ovec{x}_t$ was generated.\footnote{Note that the true label $k_t$ was assigned when the test sample $\ovec{x}_t$ was generated.  The true label $k_t$ does not necessarily indicate which of the true centroids $\{\vec{c}_k\}$ is closest to $\ovec{x}_t$.}

\begin{figure}[t!]
\centering
\begin{subfigure}[b]{\linewidth}
\centering
\psfrag{Train SSE}[][][\labelsize]{\textsf{Median SSE}}
\psfrag{M/KN}[][][\labelsize]{$M/KN$}
\includegraphics[width=\linewidth,clip]{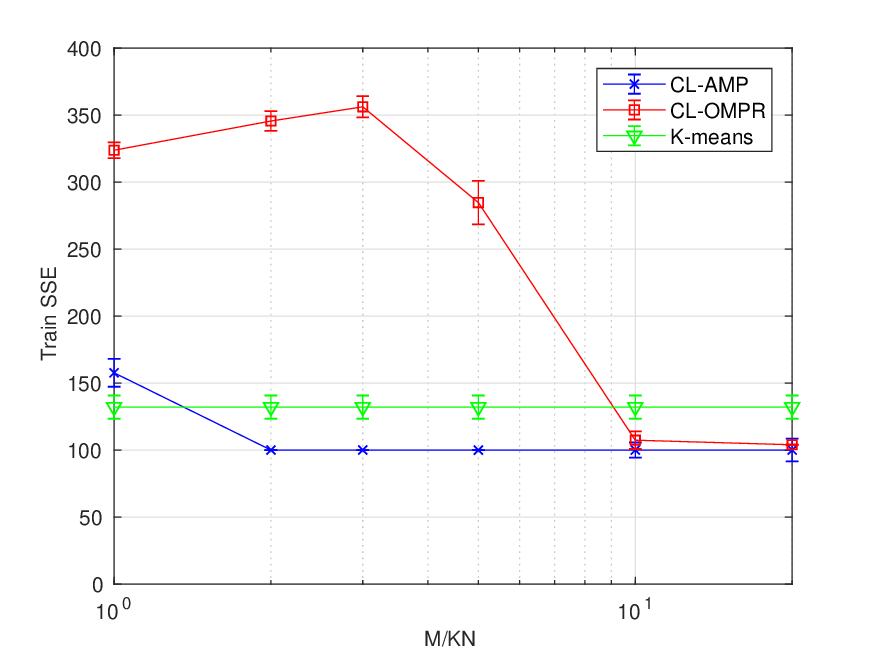}
\caption{SSE vs.\ $M$}
\label{fig:sse_v_M}
\end{subfigure}
\begin{subfigure}[b]{\linewidth}
\centering
\psfrag{Class Error}[][][\labelsize]{\textsf{Median Classification Error Rate}}
\psfrag{M/KN}[][][\labelsize]{$M/KN$}
\includegraphics[width=\linewidth,clip]{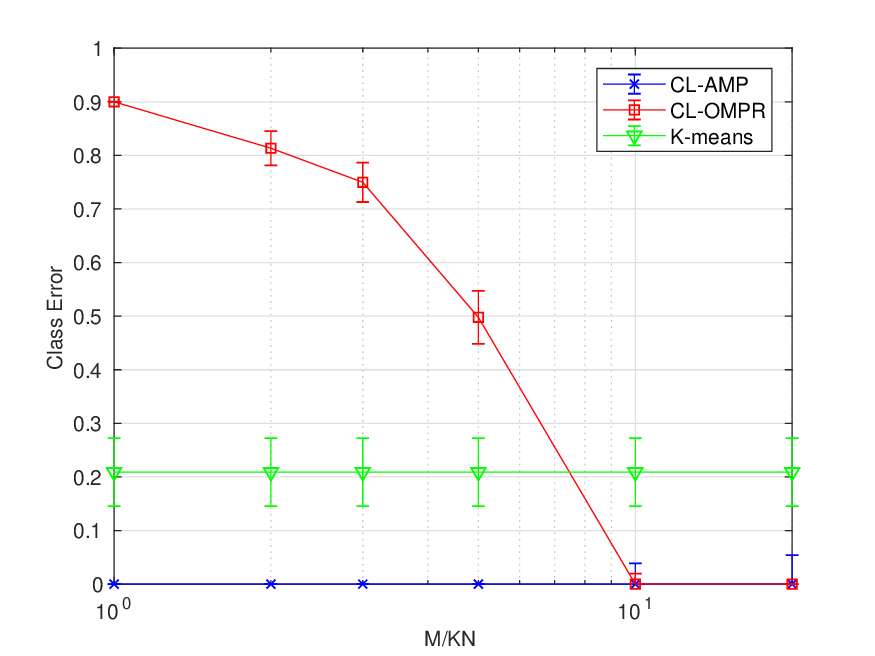}
\caption{Classification Error Rate vs.\ $M$}
\label{fig:class_v_M}
\end{subfigure}
\begin{subfigure}[b]{\linewidth}
\centering
\psfrag{Runtime w/ Sketch}[][][\labelsize]{\textsf{Median Runtime (including sketching)}}
\psfrag{M/KN}[][][\labelsize]{$M/KN$}
\includegraphics[width=\linewidth,clip]{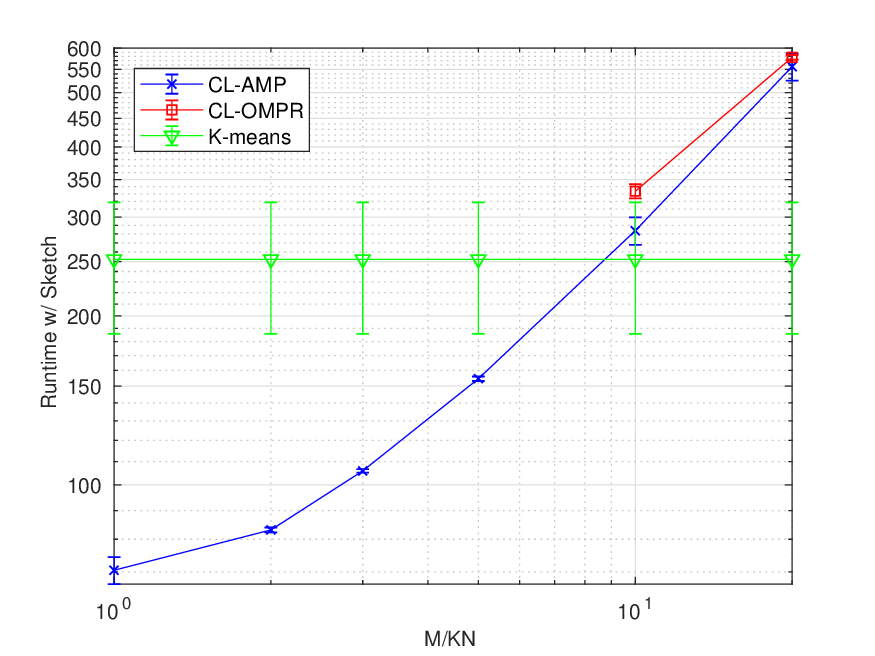}
\caption{Runtime (including sketching) vs.\ $M$}
\label{fig:time_v_M}
\end{subfigure}
\caption{Performance vs.\ sketch length $M$ for $K=10$ clusters, dimension $N=100$, and $T=10^7$ training samples.}
\label{fig:vM}
\end{figure}

Figures \ref{fig:sse_v_M}, \ref{fig:class_v_M}, and \ref{fig:time_v_M} show the median SSE, CER, and runtime (including sketching), respectively, for CL-AMP and CL-OMPR versus $M/KN$. 
Also shown is the median SSE, CER, and runtime of k-means++, as a baseline, where k-means++ has no dependence on $M$.
Because a low runtime is meaningless if the corresponding SSE is very high, the runtime was not shown for CL-AMP and CL-OMPR whenever its SSE was more than 1.5 times that of k-means++.
The error bars show the standard deviation of the estimates.

%% new figures

\Figref{sse_v_M} shows that, \textb{among the methods tested, CL-AMP achieved the lowest SSE when $M\geq 2KN$.  Also, CL-AMP supported the use of} smaller sketch sizes $M$ than CL-OMPR.
In particular, CL-AMP required $M\geq 2KN$ to yield a low SSE, while CL-OMPR required $M\geq 10KN$.
\textb{This behavior mirrors the behavior of AMP and OMP in the classical compressive sensing context, where AMP usually requires fewer measurements to accurately recover signals of a given sparsity (see, e.g., \cite[Figs. 8-10]{Vila:TSP:13}).}
Also, with sufficiently large $M$, the SSE achieved by CL-AMP and CL-OMPR was lower than that achieved by k-means++.
%Also, the minimum SSE achieved by CL-AMP was in most cases lower than that of k-means++ and CL-OMPR. 
%Among all algorithms, there was negligible difference between the Train and the Test SSE. 

\Figref{class_v_M} shows that CL-AMP achieved a low CER with sketch size $M\geq KN$, while again CL-OMPR required $M\geq 10KN$.
Also, with sufficiently large $M$, CL-AMP and CL-OMPR achieved near-zero CER, whereas k-means++ achieved an error rate of $\approx 0.2$.
%the classification error rate appears to be tied to SSE, and so similar to the SSE figure CL-AMP achieved a low classification error rate with fewer measurements than CL-OMPR, and the best classification error rate achieved by both CL-AMP and CL-OMPR was better than that of k-means++.

Finally, \figref{time_v_M} shows that, for $M/KN\in\{10,20\}$, 
k-means++ ran slightly faster than CL-AMP, 
which ran slightly faster than CL-OMPR. 
However, for $M/KN\in\{1,2,3,5\}$, CL-AMP ran significantly faster than k-means++.
For $M/KN\in\{1,2,3,5\}$, the runtime of CL-OMPR was not shown because it generated centroids of significantly worse SSE than those of k-means++.
%However, except for the smallest values of $K$ and $N$ under test, CL-AMP was faster than CL-OMPR after accounting for the improved phase transition wrt the number of measurements.
%Meanwhile, CL-AMP was approximately two order-of-magnitude slower than k-means++, although the SSE and classification error rates achieved by CL-AMP were often lower.  
%The next experiment shows a regime where CL-AMP is faster than K-means++.

\subsubsection{Performance vs.\ number of classes $K$}

\begin{figure}[t!]
\centering
\begin{subfigure}[b]{\linewidth}
\centering
\psfrag{SSE}[][][\labelsize]{\textsf{Median SSE}}
\psfrag{K}[][][\labelsize]{$K$}
\includegraphics[width=\linewidth,clip]{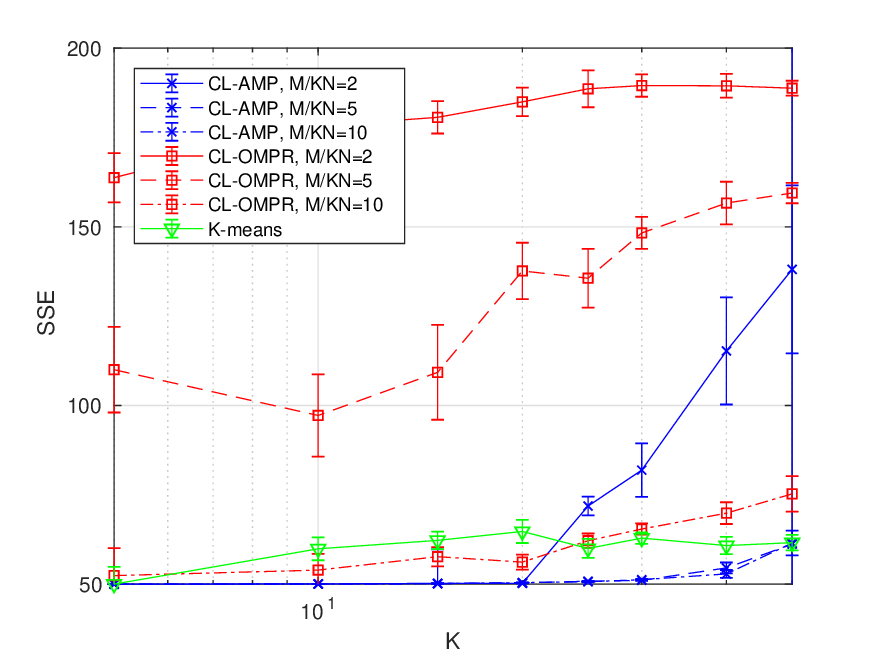}
\caption{SSE vs.\ $K$}
\label{fig:sse_v_K}
\end{subfigure}
\begin{subfigure}[b]{\linewidth}
\centering
\psfrag{Class Error Rate}[][][\labelsize]{\textsf{Median Classification Error Rate}}
\psfrag{K}[][][\labelsize]{$K$}
\includegraphics[width=\linewidth,clip]{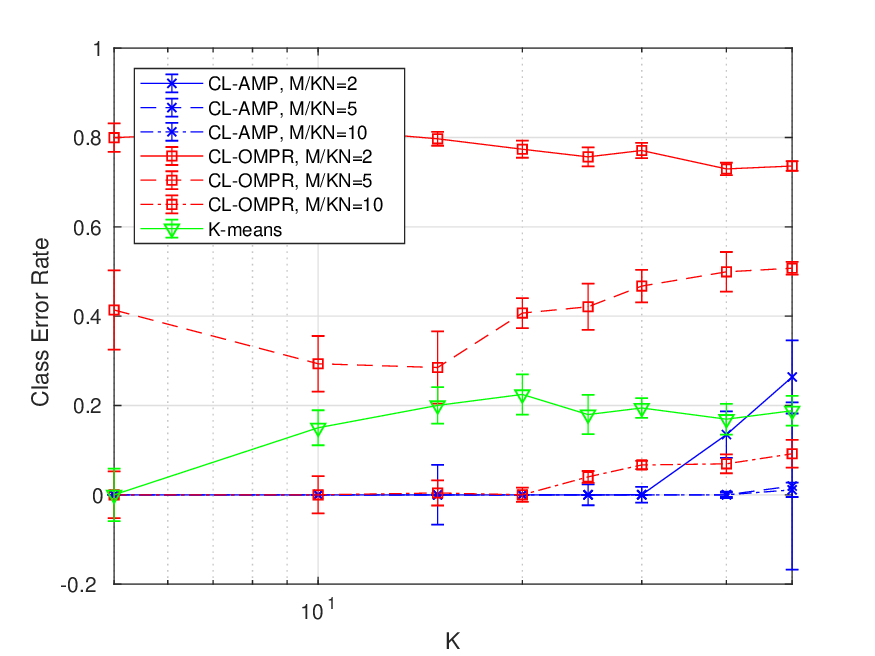}
\caption{Classification Error Rate vs.\ $K$}
\label{fig:class_v_K}
\end{subfigure}
\begin{subfigure}[b]{\linewidth}
\centering
\psfrag{Runtime w/ Sketch}[][][\labelsize]{\textsf{Median Runtime (including sketching)}}
\psfrag{K}[][][\labelsize]{$K$}
\includegraphics[width=\linewidth,clip]{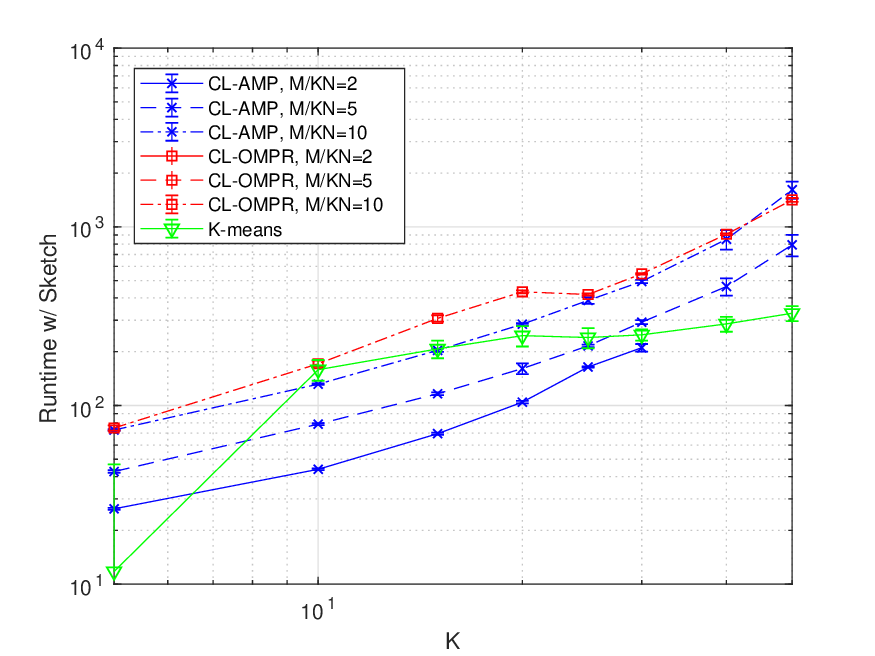}
\caption{Runtime (including sketching) vs.\ $K$}
\label{fig:time_v_K}
\end{subfigure}
\caption{Performance vs.\ number of clusters $K$ for dimension $N=50$, 
sketch size $M \in \{2,5,10\}\times KN$, and $T=10^7$ training samples.}
\label{fig:vK}
\end{figure}

In a second experiment, we evaluated each algorithm's performance versus the number of classes $K\in\{5,10,15,20,25,30,40,50\}$ and sketch sizes $M/KN\in\{2,5,10\}$ for fixed data dimension $N=50$. 
The data was generated in exactly the same way as the previous experiment, and the same performance metrics were evaluated.
%In this experiment, 10 trials were performed, where in each trial we varied $K\in\{5,10,15,20,25,30,40,50\}$ for fixed $N=50$ and $T=10^7$.
%For each trial and $K$, CL-AMP and CL-OMPR used three different sketching sizes: $M/KN\in\{2,5,10\}$.
Figures \ref{fig:sse_v_K}, \ref{fig:class_v_K}, and \ref{fig:time_v_K} show the median SSE, CER, and runtime (including sketching) versus $K$, for CL-AMP, CL-OMPR, and k-means++.

%% new figures

\Figref{sse_v_K} shows that, as $K$ increases, the SSE of k-means++ remained roughly constant, as expected based on the generation of the true centers $\vec{c}_k$.
For $K\leq 20$, CL-AMP yielded the best SSE for all tested values of $M$.
For $K>20$, CL-AMP yielded the best SSE with sketch sizes $M\in\{5KN,10KN\}$, but performed poorly with $M=2KN$. 
Meanwhile, CL-OMPR performed reasonably well with sketch size $M=10KN$, but poorly with $M\in\{2KN,5KN\}$. 

\Figref{class_v_K} shows similar trends.
With sketch size $M\in\{5KN,10KN\}$, CL-AMP had the lowest CER of any algorithm for all tested values of $K$. 
With sketch size $M=10KN$, CL-OMPR gave CER better than k-means++ for all tested $K$, but with $M\in\{2KN,5KN\}$ CL-OMPR gave CER worse than k-means++ for all tested $K$.

Finally, \figref{time_v_K} shows that CL-AMP ran faster than CL-OMPR at all tested $K$ due to its ability to work with a smaller sketch size $M$.
For large $K$, \figref{time_v_K} suggests that the runtime of both CL-AMP and CL-OMPR grow as $O(K^2)$.
The $O(K^2)$ complexity scaling is expected for CL-AMP, since its complexity is $O(MNK)$ and we set $M=O(K)$.
But the $O(K^2)$ complexity scaling is somewhat surprising for CL-OMPR, since its complexity is $O(MNK^2)$ and we set $M=10NK$.
Also, \figref{time_v_K} shows that CL-AMP ran faster than k-means++ for most values of $K$; 
for the smallest tested value of $K$ (i.e., $K=5$), the median runtime of k-means++ was lower than CL-AMP (but the error-bar suggests that the runtime of k-means++ was highly variable at this $K$).
For the largest tested value of $K$, k-means++ was again faster than CL-AMP, because the runtime of k-means++ is expected to grow linearly with $K$, whereas that of CL-AMP is expected to grow quadratically with $K$ when $M/KN$ is fixed.
%However, due to k-means++ lower complexity wrt $K$, CL-AMP was eventually slower than k-means++ for larger $K$.

\subsubsection{Performance vs.\ dimension $N$}

\begin{figure}[t!]
\centering
\begin{subfigure}[b]{\linewidth}
\centering
\psfrag{SSE/N}[][][\labelsize]{\textsf{Median SSE/N}}
\psfrag{N}[][][\labelsize]{$N$}
\includegraphics[width=\linewidth,clip]{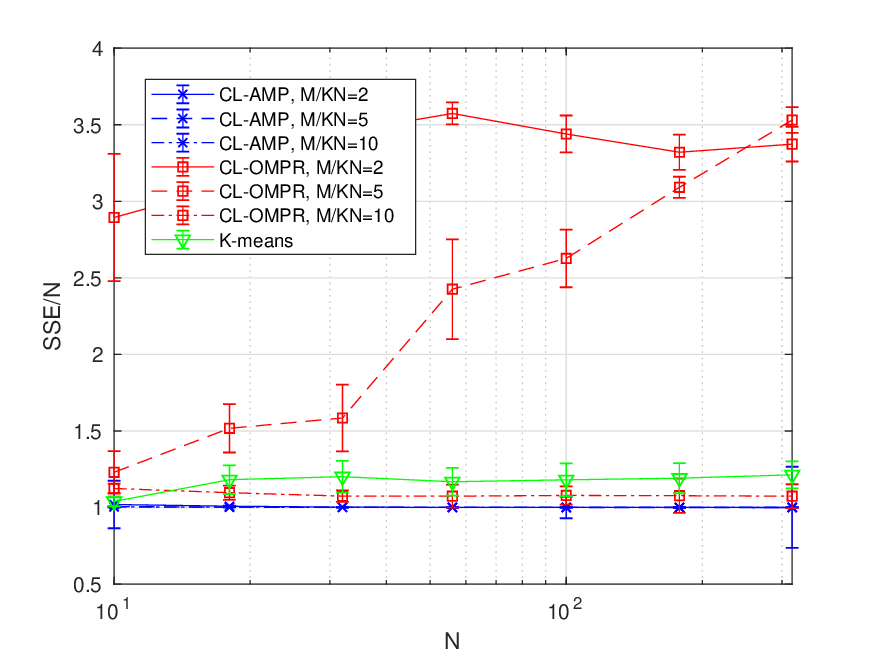}
\caption{SSE$/N$ vs.\ $N$}
\label{fig:sse_v_N}
\end{subfigure}
\begin{subfigure}[b]{\linewidth}
\centering
\psfrag{Class Error Rate}[][][\labelsize]{\textsf{Median Classification Error Rate}}
\psfrag{N}[][][\labelsize]{$N$}
\includegraphics[width=\linewidth,clip]{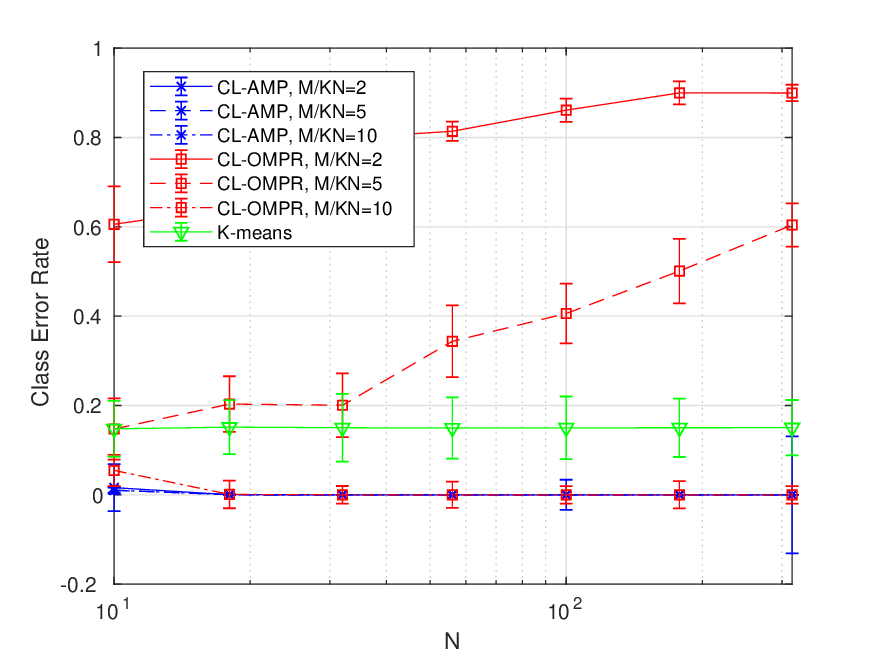}
\caption{Classification Error Rate vs.\ $N$}
\label{fig:class_v_N}
\end{subfigure}
\begin{subfigure}[b]{\linewidth}
\centering
\psfrag{Runtime w/ Sketch}[][][\labelsize]{\textsf{Median Runtime (including sketching)}}
\psfrag{N}[][][\labelsize]{$N$}
\includegraphics[width=\linewidth,clip]{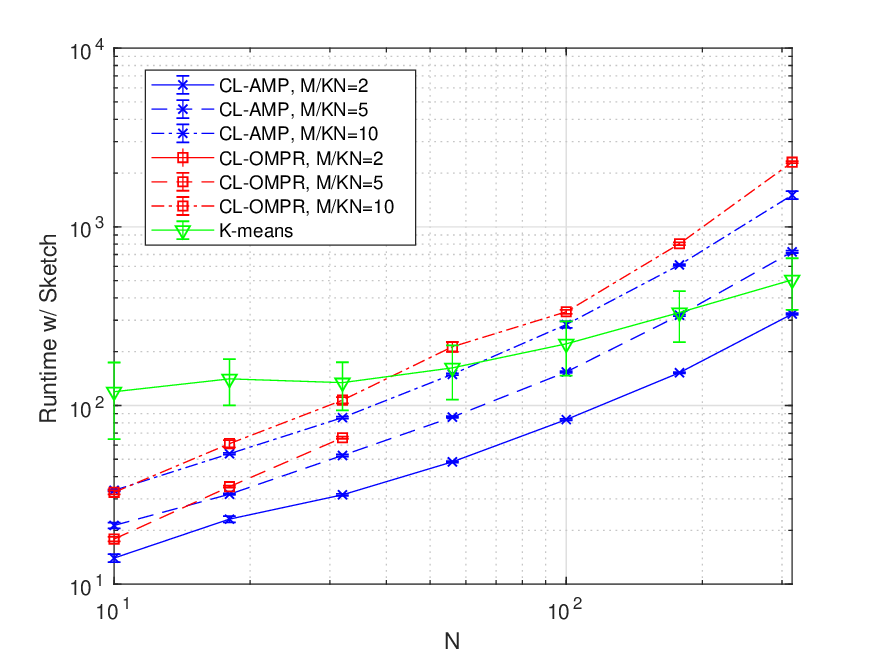}
\caption{Runtime (including sketching) vs.\ $N$}
\label{fig:time_v_N}
\end{subfigure}
\caption{Performance vs.\ dimension $N$ for $K=10$ classes, $T=10^7$ samples, and sketch size $M \in \{2,5,10\}\times KN$.}
\label{fig:vN}
\end{figure}

In a third experiment, we evaluated each algorithm's performance versus the dimension $N$ (logarithmically spaced between $10$ and $316$) for $K=10$ classes and sketch size $M\in\{2,5,10\}\times KN$.
The data was generated in exactly the same way as the previous two experiments, and the same performance metrics were evaluated.
%In this experiment, 10 trials were performed, where in each trial we varied $N$ logarithmically in $[10,316]$ for fixed $K=10$ and $T=10^7$.
%Similar to the last experiment, for each trial and $N$, CL-AMP and CL-OMPR used three different sketching sizes: $M/KN\{2,5,10\}$.
Figures \ref{fig:sse_v_N}, \ref{fig:class_v_N}, and \ref{fig:time_v_N} show the median SSE/$N$, the CER, and the runtime (including sketching) versus $N$, for CL-AMP, CL-OMPR, and k-means++.

\Figref{sse_v_N} shows that, among all algorithms, CL-AMP achieved the lowest SSE for all tested values of $N$ and $M$.
Meanwhile, both CL-OMPR under sketch size $M=10KN$ and k-means++ achieved reasonably good SSE, but CL-OMPR under smaller sketches gave much higher SSE.
%but CL-OMPR only had good SSE/$N$ when the sketch size was large, and k-means++ only had good SSE/$N$ when $N$ was small.

\Figref{class_v_N} shows that, among all algorithms, CL-AMP achieved the lowest CER for all tested values of $N$ and $M$.
Meanwhile, 
CL-OMPR under sketch size $M=10KN$ gave similar CER to CL-AMP for most $N$,
k-means++ gave significantly worse CER compared to CL-AMP for all $N$, and 
CL-OMPR under sketch size $M=5KN$ or $2KN$ gave even worse CER for all $N$.
%have only had good classification error rate when the sketch size was large, and k-means++ did not have a good classification error rate.

Finally, \figref{time_v_N} shows that, among all algorithms, CL-AMP with sketch size $M=2KN$ ran the fastest for all tested values of $N$.
Meanwhile, CL-OMPR with sketch size $M=10KN$ ran at a similar speed to CL-AMP with sketch size $M=10KN$, for all $N$. 
The runtimes for CL-OMPR with smaller sketches are not shown because it achieved significantly worse SSE than k-means++.
\Figref{time_v_N} suggests that, if $N$ is increased beyond $316$, then eventually k-means++ will be faster than CL-AMP under fixed $M/KN$.

%% new figures

\black

\subsubsection{Performance vs.\ training size $T$} \label{sec:sse_v_t}

In a final synthetic-data experiment, we evaluated each algorithm's performance versus the number of training samples $T$ (logarithmically spaced between $10^5$ and $10^8$) for $K=10$ classes, dimension $N=50$, and sketch size $M\in\{2,5,10\}KN$.
The data was generated in exactly the same way as the previous three experiments, and the same performance metrics were evaluated.
%In this experiment, we used parameters $K=10$, $N=50$, and $M/KN\in\{2,5,10\}$ and evaluated the SSE and runtime (including the time spent computing the sketch) as $T$ was varied logarithmically in $[10^5, 10^8]$.
%Everything else regarding the data generation and experimental setup was done exactly the same way as in previous experiments. 

Figures \ref{fig:sse_v_T} and \ref{fig:class_v_T} show the median SSE and CER versus $T$, for CL-AMP, CL-OMPR, and k-means++.
From these figures, we observe that the SSE and CER for each algorithm (and sketch length $M$) were approximately invariant to $T$.
CL-AMP (under any tested $M$) yielded the lowest values of SSE and CER. 
Both CL-OMPR under sketch size $M=10KN$ and k-means++ gave reasonably good SSE and CER, but CL-OMPR under smaller sketches gave worse SSE and CER.

Figures \ref{fig:time_v_T_ws} and \ref{fig:time_v_T_ns} show the median runtime with and without sketching, respectively, for the algorithms under test. 
\Figref{time_v_T_ws} shows that, if sketching time is included in runtime, then all runtimes increased linearly with training size $T$. 
However, for large $T$, CL-AMP ran faster than k-means++ and CL-OMPR (while also achieving lower SSE and CER).
Meanwhile, \figref{time_v_T_ns} shows that, if sketching time is not included in runtime, then the runtimes of both CL-AMP and CL-OMPR were relatively invariant to $T$.
Also, Figures \ref{fig:time_v_T_ws} and \ref{fig:time_v_T_ns} together show that, for $T>10^6$, the sketching time was the dominant contributer to the overall runtime.
%but meanwhile k-means++ runtime increases as $T$ becomes large.

%While not explicitly shown among the results above, CL-AMP and and CL-OMPR have more favorable memory usage than k-means (and k-means++), especially when $T$ is large.
%In particular, k-means requires storing the entire matrix $\vec{X}$, which has $NT$ elements, while CL-AMP and CL-OMPR do not store any $T$-dimensional quantities.
%Computing the sketch requires processing the $T$ data samples in $\vec{X}$, but it can be accomplished in an online manner, without storing the entire matrix in memory.

%% new figures
\begin{figure}[t!]
\centering
\begin{subfigure}[b]{\linewidth}
\centering
\psfrag{SSE}[][][\labelsize]{\textsf{Median SSE}}
\psfrag{T}[][][\labelsize]{$T$}
\includegraphics[width=\linewidth,clip]{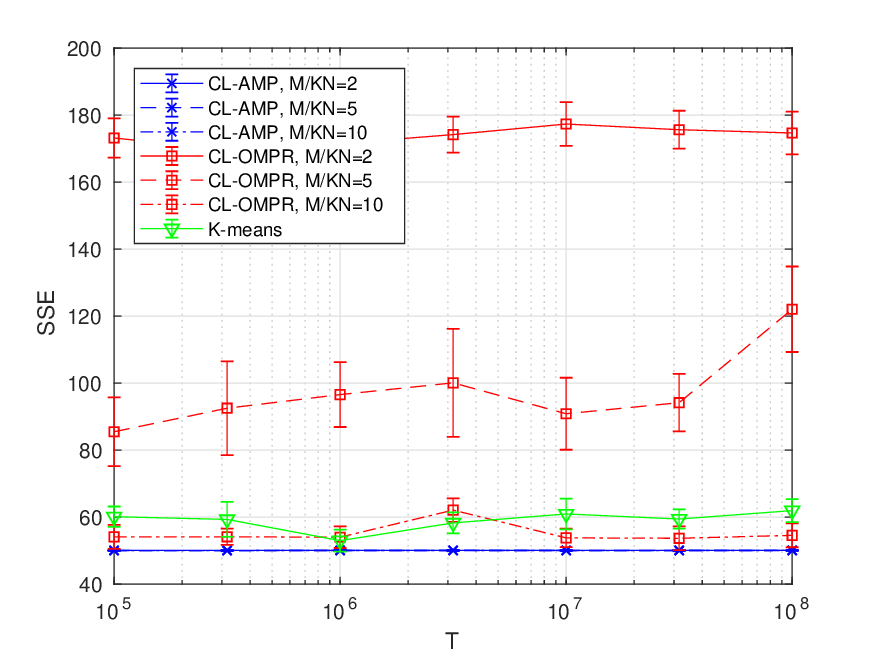}
\caption{SSE vs.\ $T$}
\label{fig:sse_v_T}
\end{subfigure}
\begin{subfigure}[b]{\linewidth}
\centering
\psfrag{Class Error Rate}[][][\labelsize]{\textsf{Median Classification Error Rate}}
\psfrag{T}[][][\labelsize]{$T$}
\includegraphics[width=\linewidth,clip]{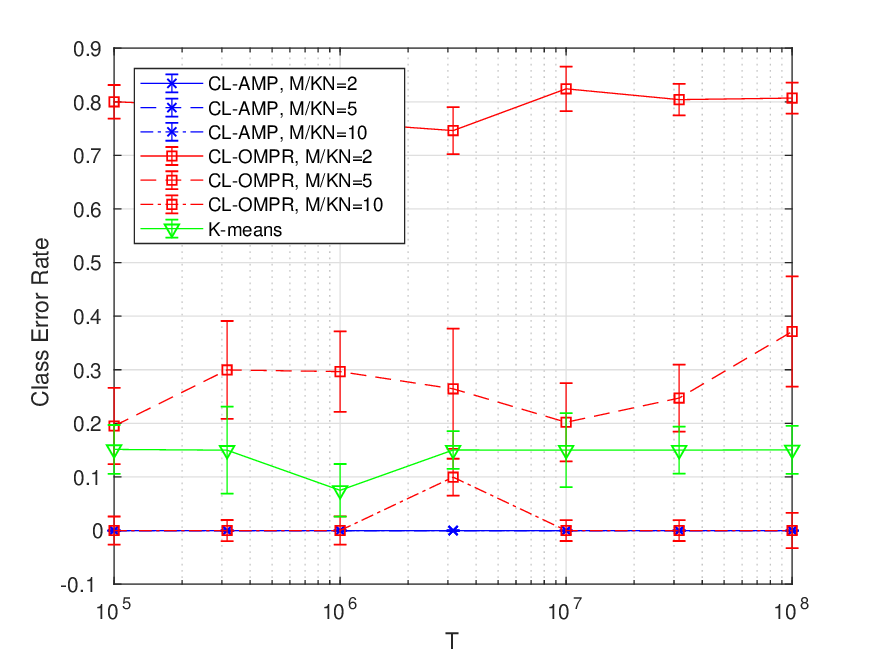}
\caption{Classification Error Rate vs.\ $T$}
\label{fig:class_v_T}
\end{subfigure}
\caption{Performance vs.\ training size $T$ for $K=10$ classes, dimension $N=50$, and sketch size $M \in \{2,5,10\}\times KN$.}
\end{figure}

\begin{figure}[t!]
\ContinuedFloat
\begin{subfigure}[b]{\linewidth}
\centering
\psfrag{Runtime w/ Sketch}[][][\labelsize]{\textsf{Median Runtime (including sketching)}}
\psfrag{T}[][][\labelsize]{$T$}
\includegraphics[width=\linewidth,clip]{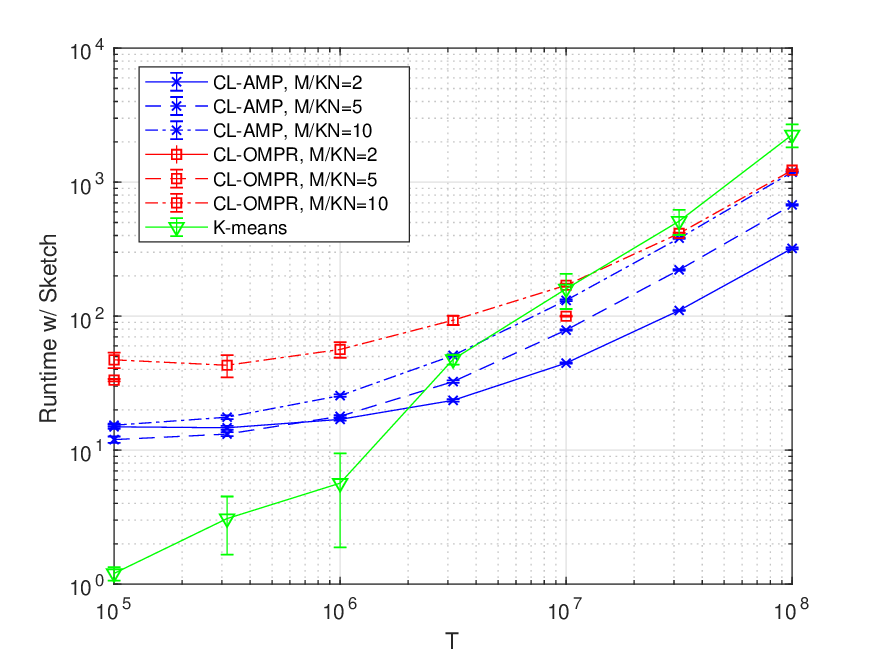}
\caption{Runtime (including sketching) vs.\ $T$}
\label{fig:time_v_T_ws}
\end{subfigure}
\begin{subfigure}[b]{\linewidth}
\centering
\psfrag{Runtime w/o Sketch}[][][\labelsize]{\textsf{Median Runtime (without sketching)}}
\psfrag{T}[][][\labelsize]{$T$}
\includegraphics[width=\linewidth,clip]{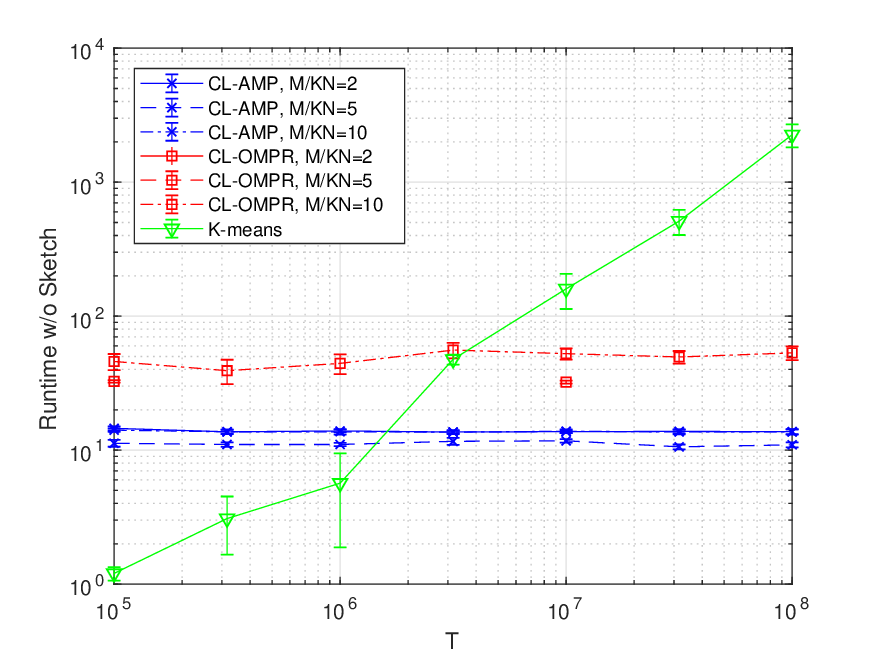}
\caption{Runtime (without sketching) vs.\ $T$}
\label{fig:time_v_T_ns}
\end{subfigure}
\caption{Performance vs.\ training size $T$ for $K=10$ classes, dimension $N=50$, and sketch size $M \in \{2,5,10\}\times KN$.}
\end{figure}

%\begin{figure}[t!]
%\centering
%\begin{subfigure}[b]{\linewidth}
%\centering
%\includegraphics[width=\linewidth,clip]{figures/sse_v_T4}
%\caption{SSE vs.\ $T$}
%\label{fig:sse_v_T}
%\end{subfigure}
%\begin{subfigure}[b]{\linewidth}
%\centering
%\includegraphics[width=\linewidth,clip]{figures/time_v_T4}
%\caption{Runtime (including sketching) vs.\ $T$}
%\label{fig:time_v_T}
%\end{subfigure}
%\caption{Performance vs.\ $T$ for  $K=10$, $N=100$, $M/KN \in \{2,10\}$.}
%\label{fig:vT}
%\end{figure}

\subsection{Spectral Clustering of MNIST} \label{sec:mnist}

Next we evaluated the algorithms on the task of spectral clustering \cite{Ng:NIPS:01} of the MNIST dataset. % \cite{LeCun:PROC:98} ??? 
This task was previously investigated for CL-OMPR and k-means++ in \cite{Keriven:ICASSP:17}, and we used the same data preprocessing steps:
extract SIFT descriptors \cite{Vedaldi:ICM:10} of each image, 
compute the $K$-nearest-neighbors adjacency matrix (for $K=10$) using FLANN \cite{Muja:VISAPP:09},
and
compute the $10$ principal eigenvectors of the associated normalized Laplacian matrix (since we know $K=10$), 
yielding features of dimension $N=10$.
%To be consistent with the CL-OMPR experiments in \cite{Keriven:ICASSP:17}, we used the preprocessed version of MNIST from \cite{Keriven:ICASSP:17}, which was preprocessed for spectral clustering \cite{Ng:NIPS:01}.
We applied this process to the original MNIST dataset, which includes $T=7\times 10^4$ samples, as well as an augmented one with $T=3\times 10^5$ samples constructed as described in \cite{Keriven:ICASSP:17}.
%The latter dataset has $K=10$ classes, dimension $N=10$, and includes two sub-datasets of size $T=7\times 10^4$ and $T=3\times 10^5$, respectively.

The experiment was conducted as follows.
In each of 10 trials, we randomly partitioned each sub-dataset into equally-sized training and testing portions.
Then, we invoked CL-AMP, CL-OMPR, and k-means++ on the training portion of the dataset, using sketch sizes $M\in\{1,2,3,5,10\}\times KN$ for CL-AMP and CL-OMPR.
The algorithm parameters were the same as in \secref{sse_min}.
Finally, the estimated centroids produced by each algorithm were evaluated using the same two metrics as in \secref{sse_min}: SSE on the training data, and classification error rate (CER) when the centroids were used for minimum-distance classification of the test data samples.
%Finally, the Adjusted Rand Index (ARI) of the training and test sets was computed.
%The ARI yields a value between less than 1, with a larger value indicating a more accurate data clustering.

\begin{figure}[t!]
\centering
\begin{subfigure}[b]{\linewidth}
\centering
\psfrag{M/KN}[][][\labelsize]{$M/KN$}
\psfrag{SSE}[][][\labelsize]{\textsf{Median SSE}}
\psfrag{SSE vs M, dataset = spec70}{}
\includegraphics[width=\linewidth,clip]{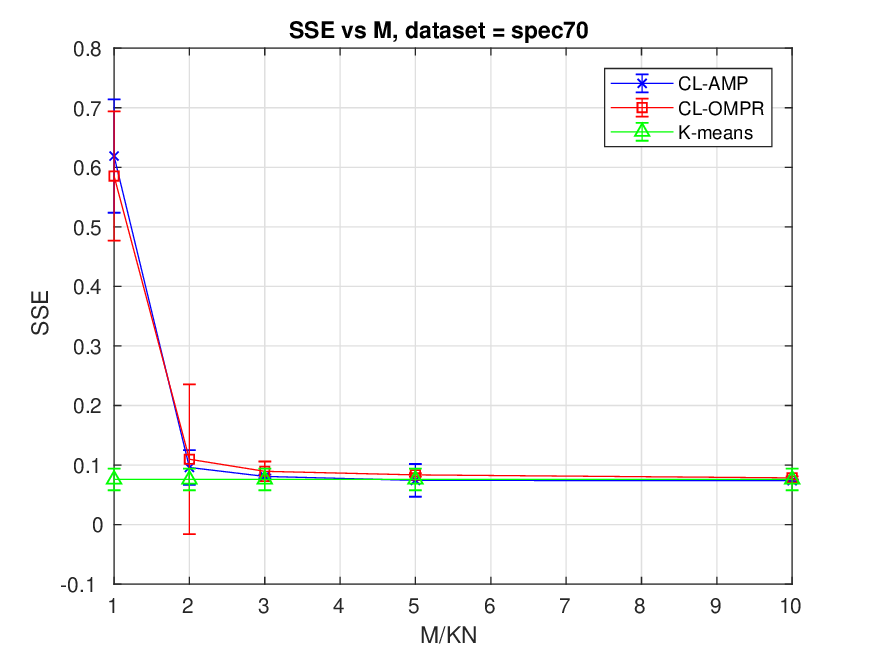}
\caption{SSE vs.\ $M$}
\label{fig:sse_v_M_mnist70}
\end{subfigure}
\begin{subfigure}[b]{\linewidth}
\centering
\psfrag{M/KN}[][][\labelsize]{$M/KN$}
\psfrag{Class Error}[][][\labelsize]{\textsf{Median Classification Error Rate}}
\psfrag{Class Error vs M, dataset = spec70}{}
\includegraphics[width=\linewidth,clip]{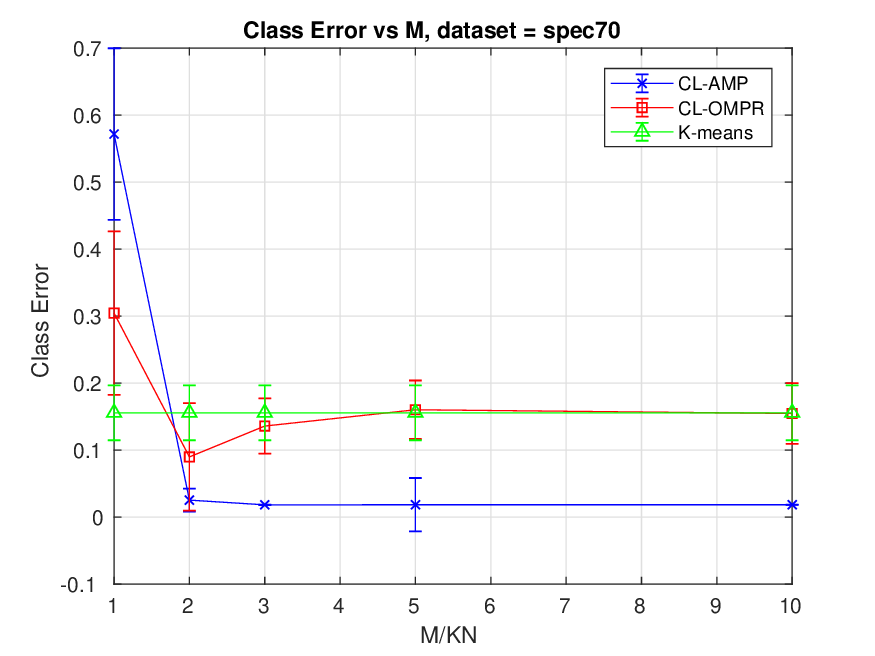}
\caption{Classification Error Rate vs.\ $M$}
\label{fig:class_v_M_mnist70}
\end{subfigure}
\begin{subfigure}[b]{\linewidth}
\centering
\psfrag{M/KN}[][][\labelsize]{$M/KN$}
\psfrag{Runtime}[][][\labelsize]{\textsf{Median Runtime (including sketching)}}
\psfrag{Time vs M, dataset = spec70}{}
\includegraphics[width=\linewidth,clip]{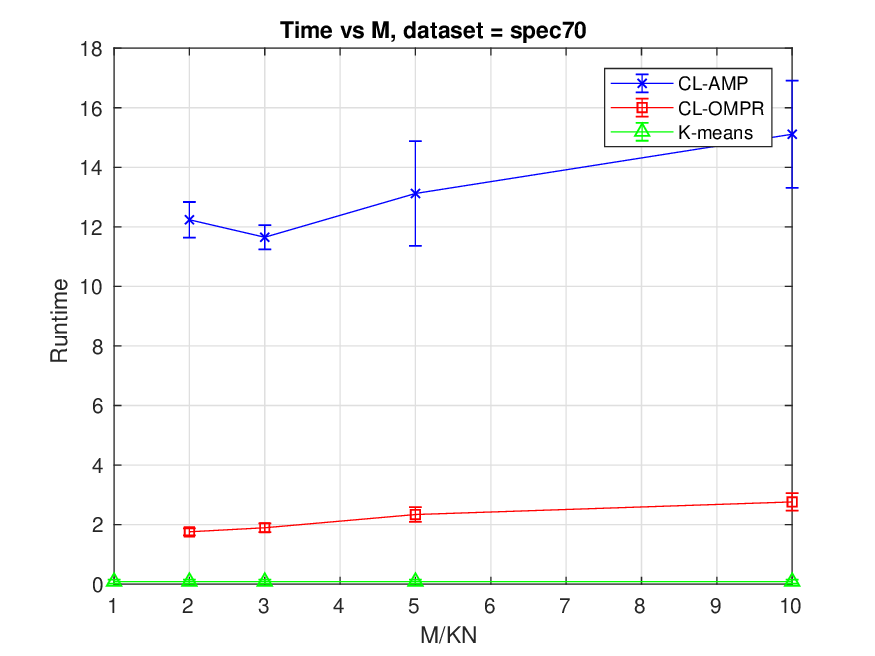}
\caption{Runtime (including sketching) vs.\ $M$}
\label{fig:time_v_M_mnist70}
\end{subfigure}
\caption{Performance vs.\ $M$ for the $T=70\,000$-sample spectral MNIST dataset, with $K=10$ clusters and dimension $N=10$.}
\label{fig:v_M_mnist70}
\end{figure}

The median SSE, CER, and runtime, versus sketch length $M$, are shown for CL-AMP and CL-OMPR in \figref{v_M_mnist70} for the $T=7\times 10^4$-sample MNIST sub-dataset. 
%and in \figref{v_M_mnist300} for $T=3\times 10^5$-sample MNIST sub-dataset. 
As before, k-means++ is shown, as a baseline, although it does not use the sketch and thus is performance is invariant to $M$.
From \textb{this figure}, we observe that CL-AMP and CL-OMPR gave respectable results for sketch lengths $M\geq 2KN$, and 
SSE nearly identical to kmeans++ for $M\geq 5KN$.
For $M\geq 2KN$, however, CL-AMP yielded significantly lower CER than both CL-OMPR and k-means++, at the cost of a slower runtime.
We attribute CL-AMP's slower runtime to its use of many iterations $i$ in \algref{tune} for hyperparameter tuning.

\blue

\subsection{Examination of Computational Complexity}

\begin{figure}[t!]
\centering
\begin{subfigure}[b]{.76\linewidth}
\centering
\psfrag{Synthetic, N=100,K=10,M/KN=2,T=1e5}{}
\includegraphics[width=\textwidth]{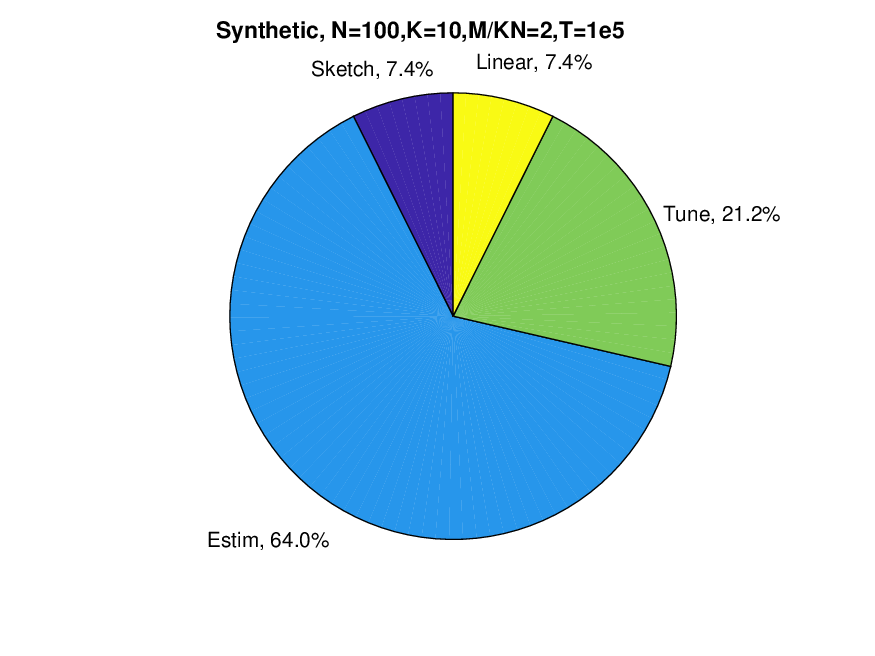}
\caption{\blue Synthetic: $N\!=\!100$, $K\!=\!10$, $M\!=\!2KN$, and $T\!=\!10^5$}
\label{fig:time1}
\end{subfigure}
\begin{subfigure}[b]{.76\linewidth}
\centering
\psfrag{Synthetic, N=100,K=10,M/KN=2,T=1e6}{}
\includegraphics[width=\textwidth]{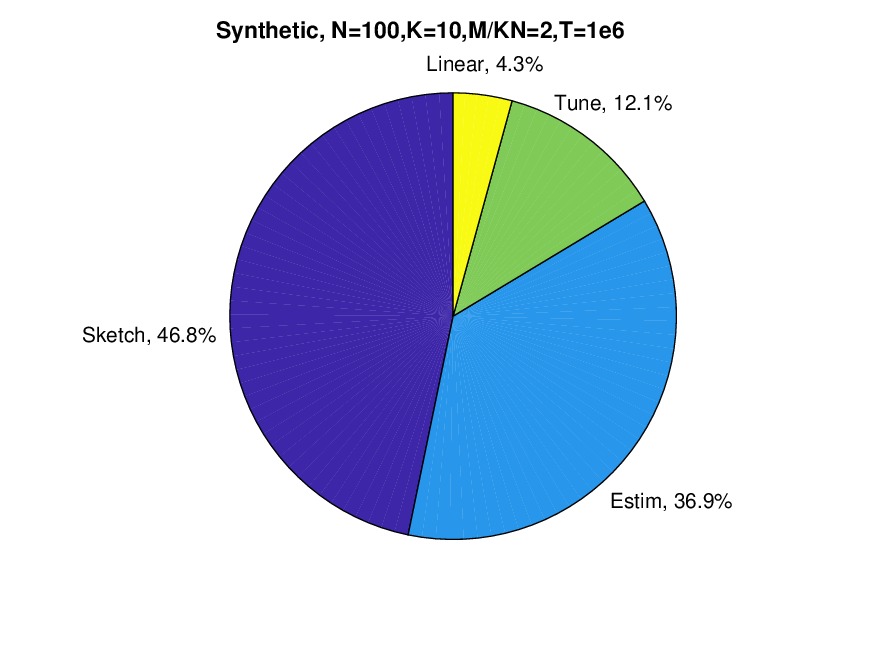}
\caption{\blue Synthetic: $N\!=\!100$, $K\!=\!10$, $M\!=\!2KN$, and $T\!=\!10^6$}
\label{fig:time2}
\end{subfigure}
\begin{subfigure}[b]{.76\linewidth}
\centering
\psfrag{Synthetic, N=100,K=10,M/KN=10,T=1e5}{}
\includegraphics[width=\textwidth]{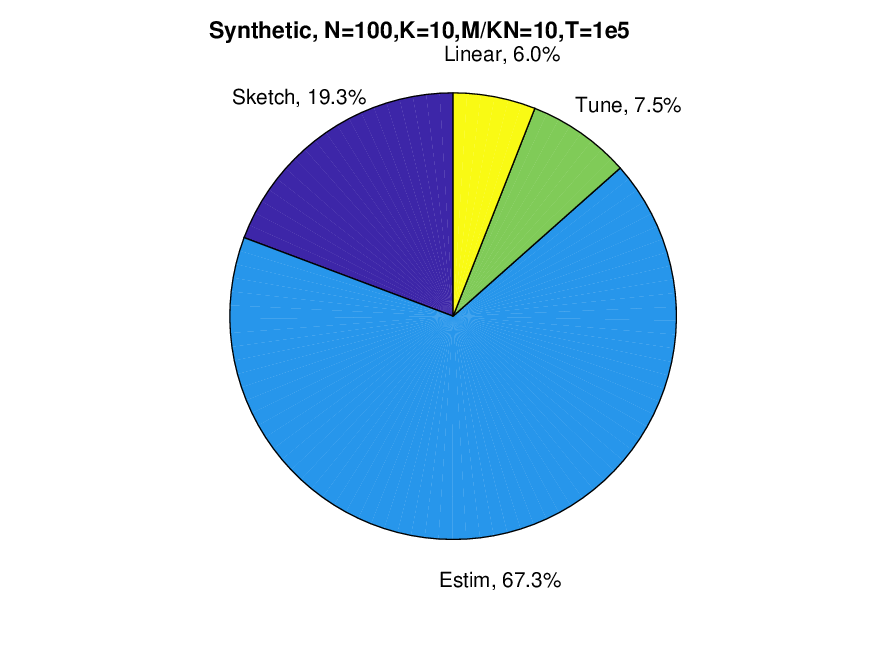}
\caption{\blue Synthetic: $N\!=\!100$, $K\!=\!10$, $M\!=\!10KN$, and $T\!=\!10^5$}
\label{fig:time3}
\end{subfigure}
\begin{subfigure}[b]{.76\linewidth}
\centering
\psfrag{MNIST, N=10,K=10,M/KN=5,T=3.5e4}{}
\includegraphics[width=\textwidth]{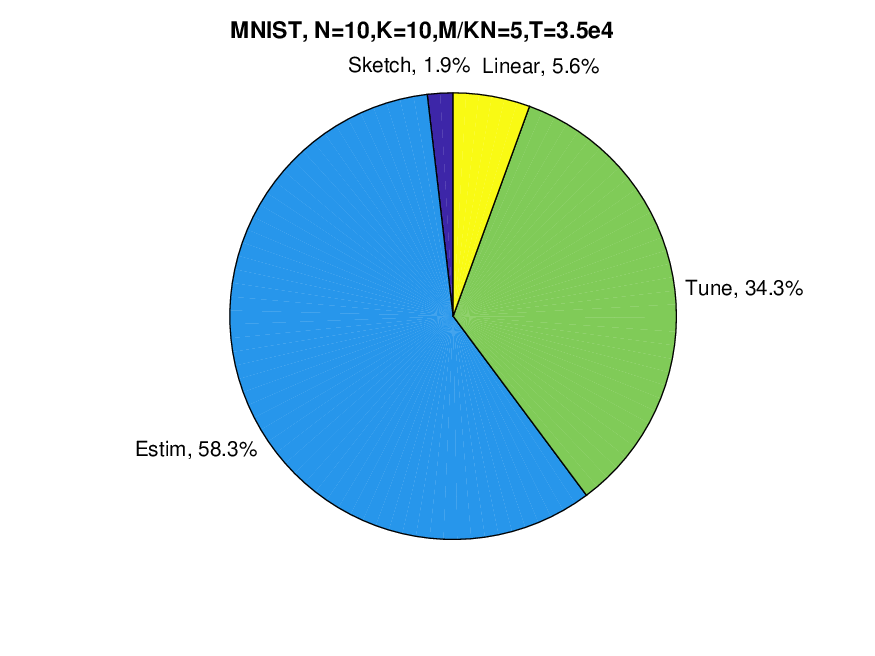}
\caption{\blue MNIST: $N\!=\!10$, $K\!=\!10$, $M\!=\!5KN$, and $T\!=\!3.5\times 10^4$}
\label{fig:time4}
\end{subfigure}
\caption{\blue Proportion of total runtime of the different sections of the CL-AMP algorithm applied to different datasets.}
\label{fig:runtimes}
\end{figure}

To better understand the computational bottlenecks of the proposed approach, 
\figref{runtimes} shows---for several problem dimensions and data types---the runtime contributions of
the ``sketch," i.e., equation \eqref{sk};
the ``tuning" steps, i.e., line~\ref{line:em} of \algref{tune};
the ``estimation" steps, i.e., lines~\ref{line:spaQz}-\ref{line:spazhat} of \algref{shygamp};
and all other lines from \algref{shygamp}, which we refer to as the ``linear" steps, since their complexity is dominated by the matrix multiplications in lines~\ref{line:phat} and \ref{line:rhat} of Algorithm 1.
%\Figref{runtimes} shows that, for the four scenarios under test, the runtime contributed by the ``linear" portion of the algorithm is relatively small.

\Figref{runtimes} suggests that CL-AMP's estimation steps require the most computation, followed by its tuning steps, and finally its linear steps.
These results motivate additional work to reduce the computational complexity of CL-AMP's estimation steps. 
The cost of sketching itself depends on the number of training samples, $T$, and the degree to which the sketching operation is distributed over multiple processors. 
When $T$ becomes large enough that the sketching time becomes computationally significant (as in \figref{time2}), the simplest remedy is to parallelize the sketch.

\black

\subsection{Frequency Estimation}

Our final experiment concerns multi-dimensional frequency estimation.
%in a different application area than clustering or classification.
Consider a sum-of-sinusoids signal of the form
\begin{equation}
\label{eq:sum_exp}
y(\vec{t}) = \sum_{k=1}^K \alpha_k \exp(\jj \vec{t}\tran \vec{c}_k), 
\end{equation}
where 
$\vec{c}_k\in\Real^N$ is the frequency of the $k$th sinusoid,
$\alpha_k>0$ is the amplitude of the $k$th sinusoid, % positive amplitudes!!!
and $\vec{t}\in\Real^N$ denotes time.
%The problem that we consider is the following:
Given measurements of the signal $y(\vec{t})$ 
at a collection of random times $\vec{t}\in\{\vec{t}_m\}_{m=1}^M$, i.e.,
\begin{equation}
y_m = y(\vec{t}_m) \text{~for~} m=1,\dots,M,
\end{equation}
we seek to recover the frequencies $\{\vec{c}_k\}_{k=1}^K$.
We are particularly interested in the case where the frequencies $\{\vec{c}_k\}$ are closely spaced, i.e., the ``super-resolution'' problem.

Note that the model in \eqref{sum_exp} matches that in \eqref{ym2} with
$g_m\vec{a}_m=\vec{t}_m ~\forall m$ and 
%\textr{$T=K$ and}
$\vec{\Phi}_k=\vec{0} \forall k$, 
%$\vec{x}_t = \vec{c}_t$, and $\tau_{mk} = 0$, 
so that we can apply CL-AMP to this frequency estimation problem.
The model in \eqref{sum_exp} also matches \eqref{sk_clompr} with
$\vec{w}_m=\vec{t}_m~\forall m$, and so we can also apply CL-OMPR.
But we cannot apply k-means++.

For frequency pairs $\{\vec{c}_1, \vec{c}_2\}$ with $\norm{\vec{c}_1-\vec{c}_2}_2 \geq \epsilon$, 
\cite{Traonmilin:SPARS:17} claims that, with $\{\vec{w}_m\}$ drawn randomly from an appropriate distribution, one can resolve the frequencies with $M\geq O\big(\ln(1/\epsilon)\big)$ measurements. 
However, choosing $\vec{w}_m$ uniformly spaced on a grid would require $M\geq O(1/\epsilon)$ measurements.
Thus, for a final experiment, similar to those performed in \cite{Traonmilin:SPARS:17}, we did the following.
For a particular $N$ and $K$ (where $K$ is even for simplicity), we generated $K/2$ pairs of frequencies $\{\vec{c}_{2k-1},\vec{c}_{2k}\}$, where $\norm{\vec{c}_{2k-1} - \vec{c}_{2k}}_2 = \epsilon$ for $k=1,...,K/2$.
Then, for a particular realization of $\{\vec{c}_k\}_{k=1}^K$ and $\{\vec{w}_m\}_{m=1}^M$, CL-AMP and CL-OMPR were invoked to estimate $\{\hvec{c}_k\}_{k=1}^K$.
Recovery was declared successful if 
\begin{equation}
%\sqrt{\min_{\{j_1,...j_K\}=\{1,...,K\}}  \frac{1}{K}\sum_{k=1}^K \norm{\vec{c}_{j_k} - \hvec{c}_{k}}_2^2} < \frac{\epsilon}{3}.
%\max_k \bigg\{\min_{\{j_1,...j_K\}=\{1,...,K\}}  \norm{\vec{c}_{j_k} - \hvec{c}_{k}}_2 \bigg\} < \frac{\epsilon}{2},
\max_k \norm{\vec{c}_{j_k} - \hvec{c}_{k}}_2 < \epsilon/2,
\end{equation}
where $\{j_k\}_{k=1}^K$ solves the linear assignment problem \eqref{LAP}.
%i.e., there exists an assignment of the estimated centroids for which none is more that $\epsilon/2$ from the true centroid.

\begin{figure}[t!]
\centering
\begin{subfigure}[b]{\linewidth}
\centering
\psfrag{SC-AMP; K = 4, N = 2, draw = fg, trials = 10}{}
\psfrag{M/(KN)}[][][\labelsize]{$M/KN$}
\psfrag{.1/e}[l][l][0.6]{$.1/\epsilon$}
\psfrag{log(1/e)}[l][l][0.65]{$\ln(1/\epsilon)$}
\includegraphics[width=\linewidth,clip]{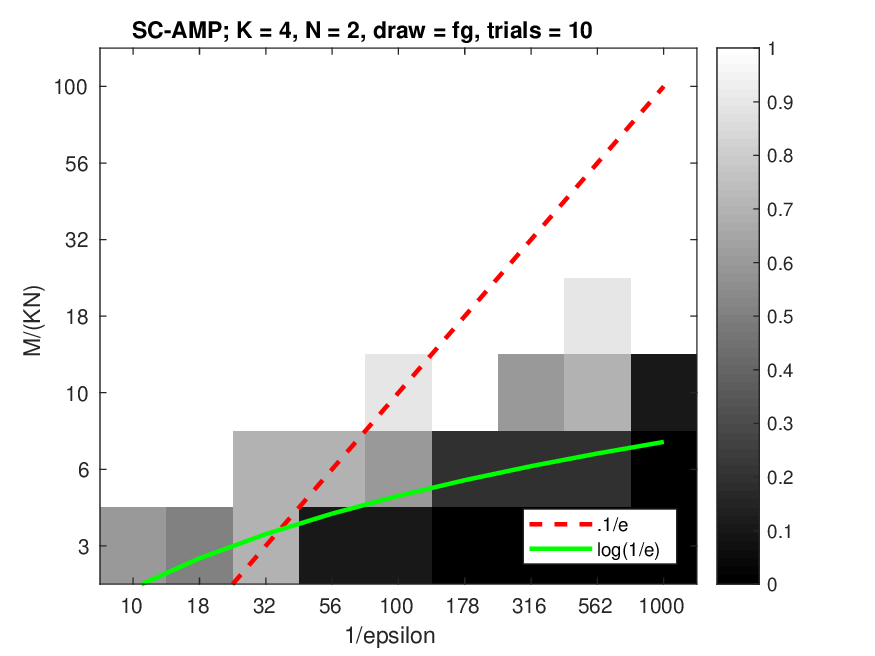}
\caption{CL-AMP}
\label{fig:clamp_k4}
\end{subfigure}
\begin{subfigure}[b]{\linewidth}
\centering
\psfrag{CL-OMPR; K = 4, N = 2, draw = fg, trials = 10}{}
\psfrag{M/(KN)}[][][\labelsize]{$M/KN$}
\psfrag{.1/e}[l][l][0.6]{$.1/\epsilon$}
\psfrag{log(1/e)}[l][l][0.65]{$\ln(1/\epsilon)$}
\includegraphics[width=\linewidth,clip]{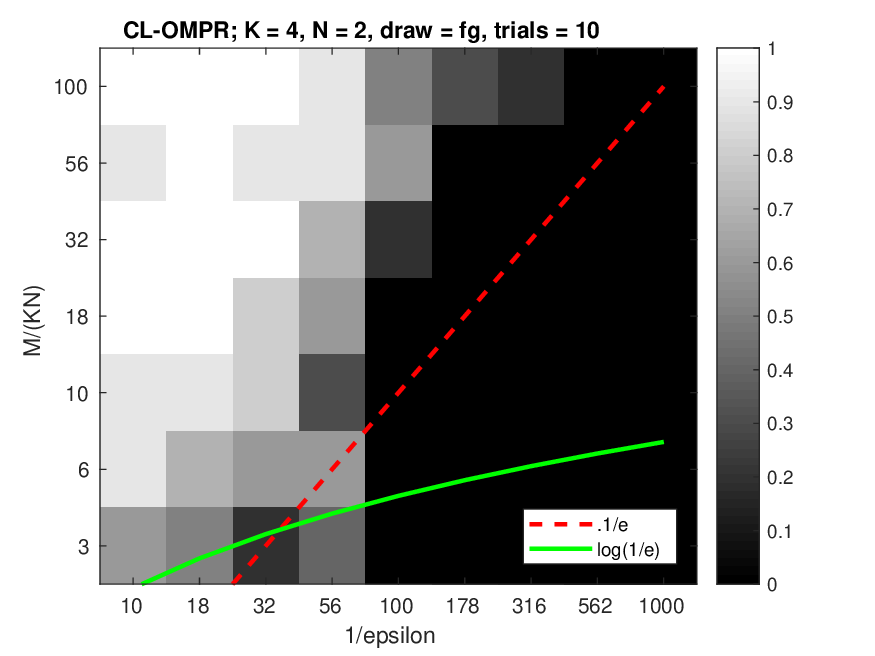}
\caption{CL-OMPR}
\label{fig:clompr_k4}
\end{subfigure}
\caption{Frequency estimation for $K=4$ and $N=2$ with random time samples.}
\label{fig:freq_estim4}
\end{figure}

\begin{figure}[t!]
\centering
\begin{subfigure}[b]{\linewidth}
\centering
\psfrag{SC-AMP; K = 4, N = 2, draw = unif, trials = 10}{}
\psfrag{M/(KN)}[][][\labelsize]{$M/KN$}
\psfrag{.1/e}[l][l][0.6]{$.1/\epsilon$}
\psfrag{log(1/e)}[l][l][0.65]{$\ln(1/\epsilon)$}
\includegraphics[width=\linewidth,clip]{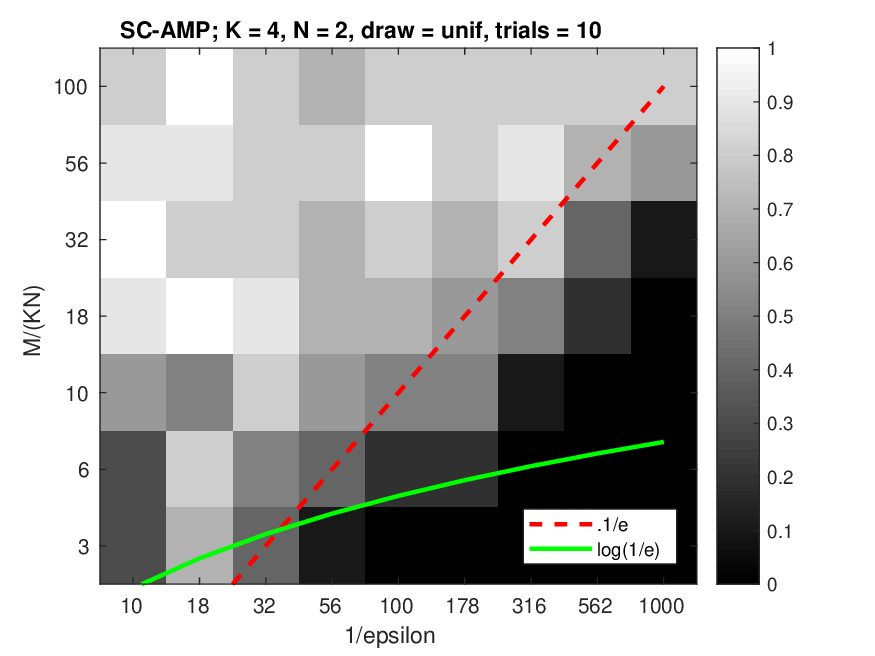}
\caption{CL-AMP}
\label{fig:clamp_k4_unif}
\end{subfigure}
\begin{subfigure}[b]{\linewidth}
\centering
\psfrag{CL-OMPR; K = 4, N = 2, draw = unif, trials = 10}{}
\psfrag{M/(KN)}[][][\labelsize]{$M/KN$}
\psfrag{.1/e}[l][l][0.6]{$.1/\epsilon$}
\psfrag{log(1/e)}[l][l][0.65]{$\ln(1/\epsilon)$}
\includegraphics[width=\linewidth,clip]{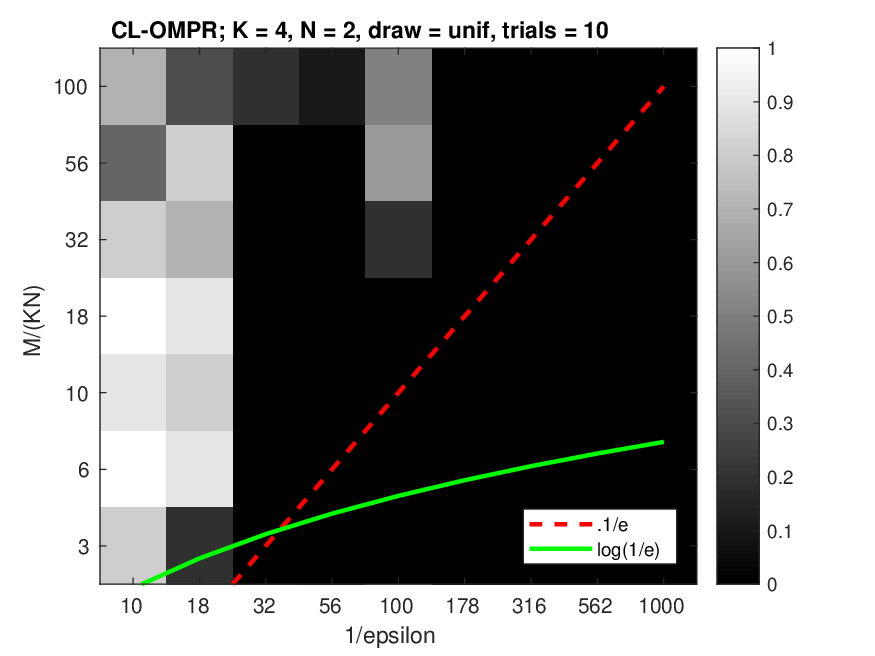}
\caption{CL-OMPR}
\label{fig:clompr_k4_unif}
\end{subfigure}
\caption{Frequency estimation for $K=4$ and $N=2$ with uniformly spaced time samples.}
\label{fig:freq_estim4_unif}
\end{figure}

For our experiment, we tested $K\!=\!4$ frequency components of dimension $N\!=\!2$ and varied $M$ from $3KN$ to $100KN$ while also varying $\epsilon$ from $10^{-1}$ to $10^{-3}$.
For each combination, 10 trials were performed. 
The empirical probability of successful recovery is shown in Figures~\ref{fig:freq_estim4}-\ref{fig:freq_estim4_unif}.
In \figref{freq_estim4}, $\vec{a}_m$ were drawn uniformly on the unit sphere and $g_m\!=\!|g_m'|$ with $g_m'\!\sim\!\mc{N}\big(0,4\epsilon^2 \log_{10}^2(\epsilon)\big)$, \textb{while in \figref{freq_estim4_unif}, $\vec{w}_m$ are uniformly spaced on a $N$-dimensional hyper-grid with per-dimension spacing $\pi/2$}. 
Superimposed on the figures are curves showing $M/KN = 0.1/\epsilon$ and $M/KN = \ln(1/\epsilon)$.
From the figures, we see that CL-AMP had a higher empirical probability of recovery than CL-OMPR, especially for small $\epsilon$.
%We also see that the recovery performance of both algorithms was superior with random time samples versus uniformly spaced time samples.
We also see that the empirical phase transition of CL-AMP is close to the $\ln(1/\epsilon)$ curve with random frequency samples \textb{(i.e., \figref{clamp_k4})}
and the $0.1/\epsilon$ curve with uniform frequency samples \textb{(i.e., \figref{clamp_k4_unif})}.

%-------------------------------------------------------------------------------
\section{Conclusion} \label{sec:conc}

In sketched clustering, the original dataset is sketched down to a relatively short vector, from which the centroids are extracted.
For the sketch proposed by \cite{Keriven:II:17,Keriven:ICASSP:17}, we proposed the ``CL-AMP'' centroid-extraction method. 
Our method assumes that the original data follows a GMM, and exploits the recently proposed simplified hybrid generalized approximate message passing (SHyGAMP) algorithm \cite{Byrne:TSP:16}.
%In our case we assumed a Gaussian Mixture Model on our original dataset, which when combined with our sketching operator, yielded a generalized linear model parameterized by the true cluster centers.
%In order to estimate the cluster centers under this model, we invoked the Simplified-Hybrid-GAMP algorithm, yielding the CL-AMP algorithm.
Numerical experiments suggest that CL-AMP exhibits better sample complexity (i.e., extracts accurate clusters with fewer compressed samples) than the state-of-the-art sketched-clustering algorithm, CL-OMPR, from \cite{Keriven:II:17,Keriven:ICASSP:17}.
In many cases, CL-AMP also exhibits better computational complexity than CL-OMPR.
Furthermore, for datasets with many samples, CL-AMP exhibits lower computational complexity than the widely used k-means++ algorithm.
%Through several numerical experiments, we have shown that our approach to sketched clustering is faster than the existing state-of-the-art sketched clustering algorithm CL-OMPR \cite{Keriven:II:17,Keriven:ICASSP:17} in high dimensions, as well as in sample complexity (i.e., CL-AMP obtains better estimates than CL-OMPR for a given size of the sketch).
%Additionally, in some regimes where the original dataset is extremely large, CL-AMP if more efficient than the widely used k-means++ algorithm.
As future work, 
\textb{
it would be worthwhile to investigate ways to reduce the computational complexity of CL-AMP's estimation steps, and to analyze the theoretical behavior of CL-AMP using a state-evolution approach.
%Or, when the number of training samples $T$ is large, in which case the complexity of the sketch in \eqref{sk} dominates the complexity of CL-AMP, it may be worthwhile to investigate ways to reduce the complexity of the sketch.
%(e.g., \cite{Chatalic:ICASSP:18}).
Finally, as new variations of the sketch \eqref{sk} are proposed (e.g., \cite{Chatalic:ICASSP:18,Schellekens:SPL:18}) it would be interesting to modify CL-AMP accordingly.
}

\bibliographystyle{IEEEtran}
\bibliography{macros_abbrev,stc,books,misc,sparse,machine,phase}

\end{document}